# IMPLEMENTATION OF THE OPEN SOURCE VIRTUALIZATION TECHNOLOGIES IN CLOUD COMPUTING

Mohammad Mamun Or Rashid, M. Masud Rana and Jugal Krishna Das

Department of Computer Science and Engineering, Jahangirnagar University Savar, Dhaka, Bangladesh
fedoraboy@gmail.com, masudrana.ppi@gmail.com and cedas@juniv.edu

## *ABSTRACT*

*The "Virtualization and Cloud Computing" is a recent buzzword in the digital world. Behind this fancy poetic phrase there lies a true picture of future computing for both in technical and social perspective. Though the "Virtualization and Cloud Computing are recent but the idea of centralizing computation and storage in distributed data centres maintained by any third party companies is not new but it came in way back in 1990s along with distributed computing approaches like grid computing, Clustering and Network load Balancing. Cloud computing provide IT as a service to the users on-demand basis. This service has greater flexibility, availability, reliability and scalability with utility computing model. This new concept of computing has an immense potential in it to be used in the field of e-governance and in the overall IT development perspective in developing countries like Bangladesh.*

## 1. INTRODUCTION

Administrators usually use lots of servers with heavy hardware to keep their service accessible, available for the authenticated users. As days passes by concernment of new services increases which require more hardware, more effort from IT administrators. There is another issue of capacity (Hardware as well as storage and networking) which always increases day by day. Moreover sometime we need to upgrade old running servers as their resources have been occupied fully. On that case we need to buy new servers, install those services on that server and finally migrate to the service on it. Cloud computing focus on what IT always needs: a way to increase capacity on the fly without investing in new infrastructure. Cloud computing also encompasses any subscription-based, user-based, services-based or pay-per-use service that in real time over the internet extends its existing capabilities.

### 1.1 Definition of Cloud Computing

Cloud computing is a model for enabling convenient, on-demand network access to a shared pool of configurable computing resources (e.g., networks, servers, storage, applications and services) that can be rapidly provisioned and released with minimal management effort on service provider interaction [1].

### 1.2 Benefits of Cloud Computing

**Flexibility** – Every day organization demands increase and Scale up or down to meet their requirements. Today's economy, flexibility is the key. One can adjust his IT expenditures to meet your organization's immediate needs.

**Security –** Cloud service assured that your data in the cloud is much more secure than in your small unsecured server room.

**Capacity** –With cloud computing, Capacity always increase and it is no longer an issue. Now, focus on how the solution will help in further mission. The IT piece belongs to somebody else.

**Cost** – Cloud and Virtualization technology reduce your all maintenance fees. There is no more servers, software, and update fees. Many of the hidden costs typically associated with software implementation, customization, hardware, maintenance, and training are rolled into a transparent subscription fee.

## 1.3 Virtualization

Virtualization can be practical very broadly to just about everything you can imagine including processor, memory, networks, storage, operating systems, and applications. Three characteristics of virtualization technology make it ideal for cloud computing:

**Partitioning:** In virtualization technology, single physical server or system can use partitioning to support many different applications and operating systems (OS).

**Isolation:** In cloud computing, each virtual machine is isolated and protected from crashes or viruses in the other machines. What makes virtualization so important for the cloud is that it decouples the software from the hardware.

**Encapsulation:** Encapsulation can protect each application so that it doesn't interfere with other applications. By using encapsulation, each virtual machine stored as a single file, making it easy to identify and present to other applications and software. To understand how virtualization helps with cloud computing, we must understand its many forms. In all cases, a single resource actually emulates or imitates other resources. Here are some examples:

**Virtual memory:** Every disk has a lot more space than memory. PCs can use virtual memory to borrow extra memory from the hard disk. Although virtual disks are slower than real memory, if managed right, the substitution works surprisingly well.

**Software:** Virtualization software is available which can emulate an entire computer. A virtual single computer can perform as though it were actually more than computers. This kind of software might be able to move from a data centre with thousands of servers. To manage virtualization in cloud computing, most of companies are using different hypervisors. Because in cloud computing we need different operating environments, the hypervisor becomes an ideal delivery mechanism by allowing same application on lots of different systems. Hypervisors can load multiple operating systems in single node; they are a very practical way of getting things virtualized quickly and efficiently. Let's try to draw a picture on above statement.

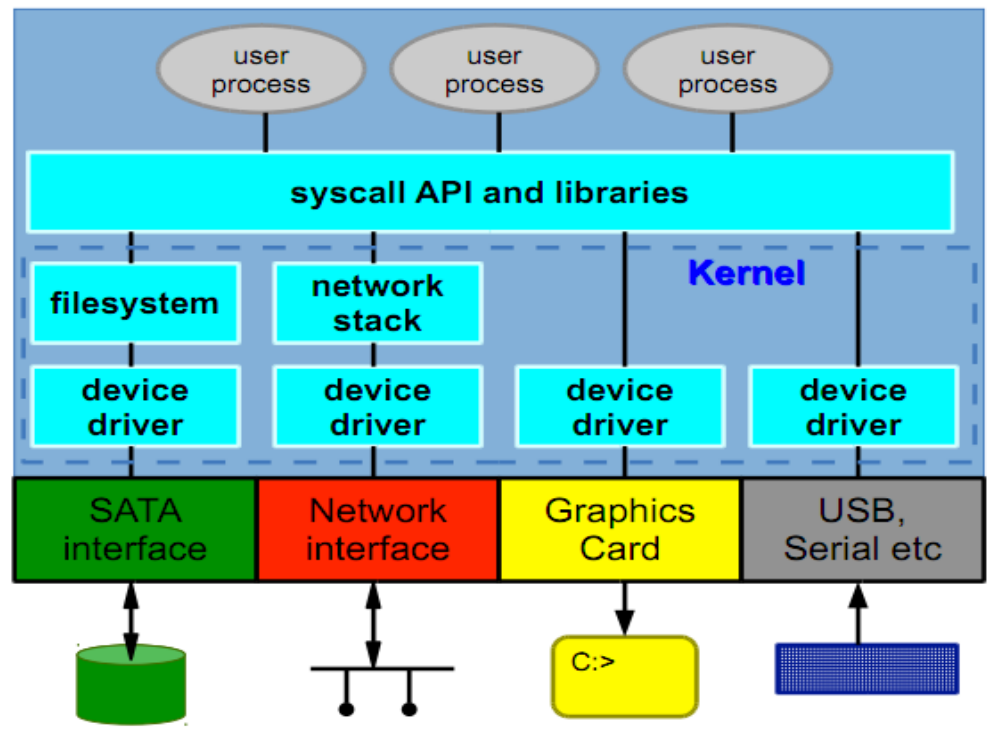

Figure 1.1: A normal Workstation / Computer

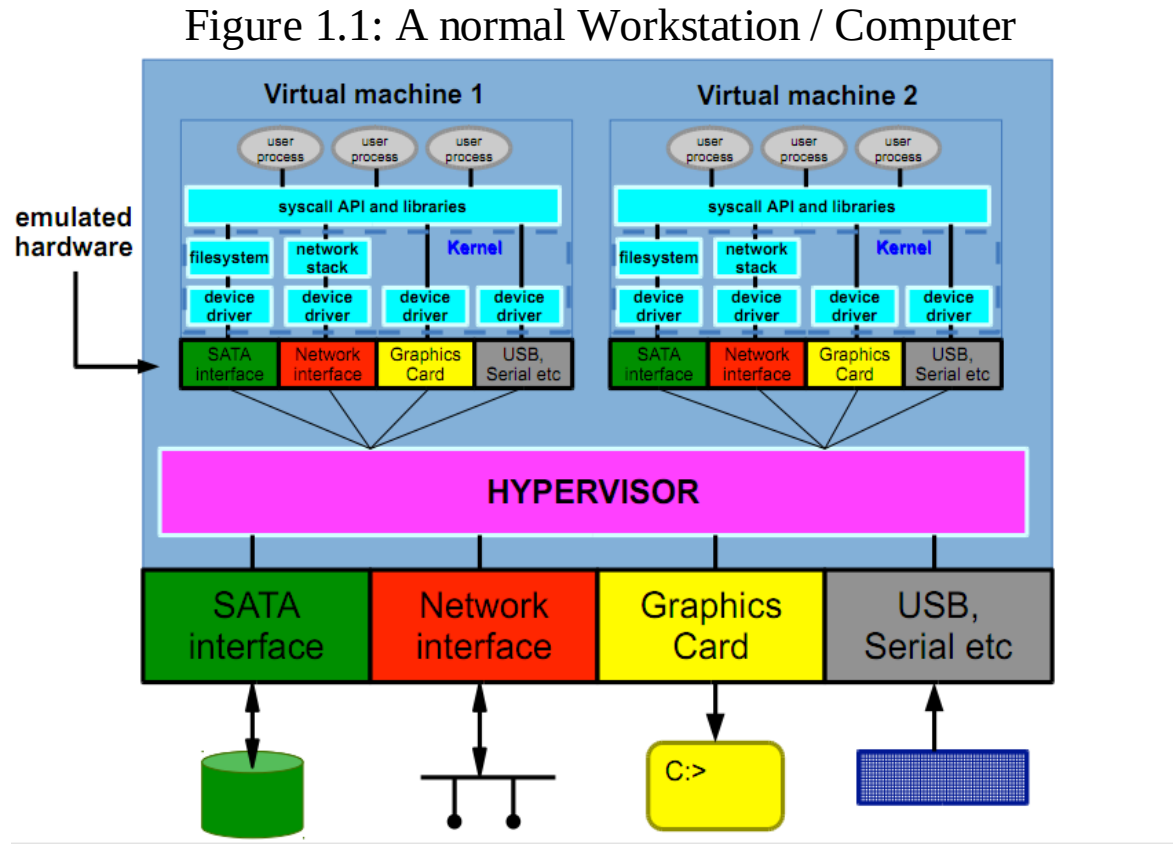

Figure 1.2: Workstation using Hypervisor

## 1.4 Hypervisor

The evolution of virtualization greatly revolves around one piece of very important software that loads the whole virtual system. This is the hypervisor. As an integral component of computer node, this software allows all physical devices to share their all resources (Processor, RAM, Disk, Network) amongst virtual machines running as guests on to top of that physical hardware devices.

## 1.5 Related Work

Open source Virtualization technologies in Cloud computing provided this paper on multiple Node to measure its performance [2], [3], [4] and [5]. In this paper, we extend this evaluation to include Master Node as another Instance in virtualization platform, and test both under different scenarios including multiple VMs and multi-tiered systems. We have also working with oVirt in Virtualization that implemented with Centos 6. We created three Hypervisor (Node) and One Manager. There are 76 Virtual Machine running where most of them application Server and 4 Database server with Disaster Recovery System. For Application server, We have implemented NBL(Network Load Balancer) for web services to active in service 24/7. Ganeti supports a very lightweight architecture which is very useful to start with commodity hardware. From starting a single node installation an administrator can scale out the cluster very easily. It is designed to use local storage also compatible with larger storage solutions. It has fault-tolerance as a built-in feature. In a word it is very simple to manage and maintain. Ganeti is admin centric clustering solution which is the main barrier for public cloud deployment. To the best of our knowledge, these types of virtualization technologies have not been evaluated in the context of server clustering. Multiple Node Server consolidation using virtual containers brings new challenges and, we comprehensively evaluate two representative virtualization and cloud technologies in a number of different Node scenarios.

# IMPLEMENTATION

## 2.1 Scope of this project

In this project we used following configuration hardware.

CPU: Dual Core
RAM: 2GB
Storage: 140GB
NIC: 1

We use 3 hardwires stated like above. We will use Debian GNU/Linux 7 as our base operating system and run Ganeti over the operating system using KVM as hypervisor. Later we will initiate a cluster on one physical host as a master node. We will join other nodes on that cluster. We will use a manageable switch and VLAN on it to separate our management + storage network and public facing VM network for security purpose. Later we will create VMs and check live migration, Network changes and failover scenarios.

## 2.2 Summary of the topology

We will connect three commodity computers in our cluster. Each computer has a single NIC which will be logically divided by VLANs. All the computers will be connected to a trunk port of a manageable switch to accept logical network (VLAN). The management + Storage network and pubic network (VM) will be separated from that manageable switch. The deployment architecture and physical node connectivity has been presented below.

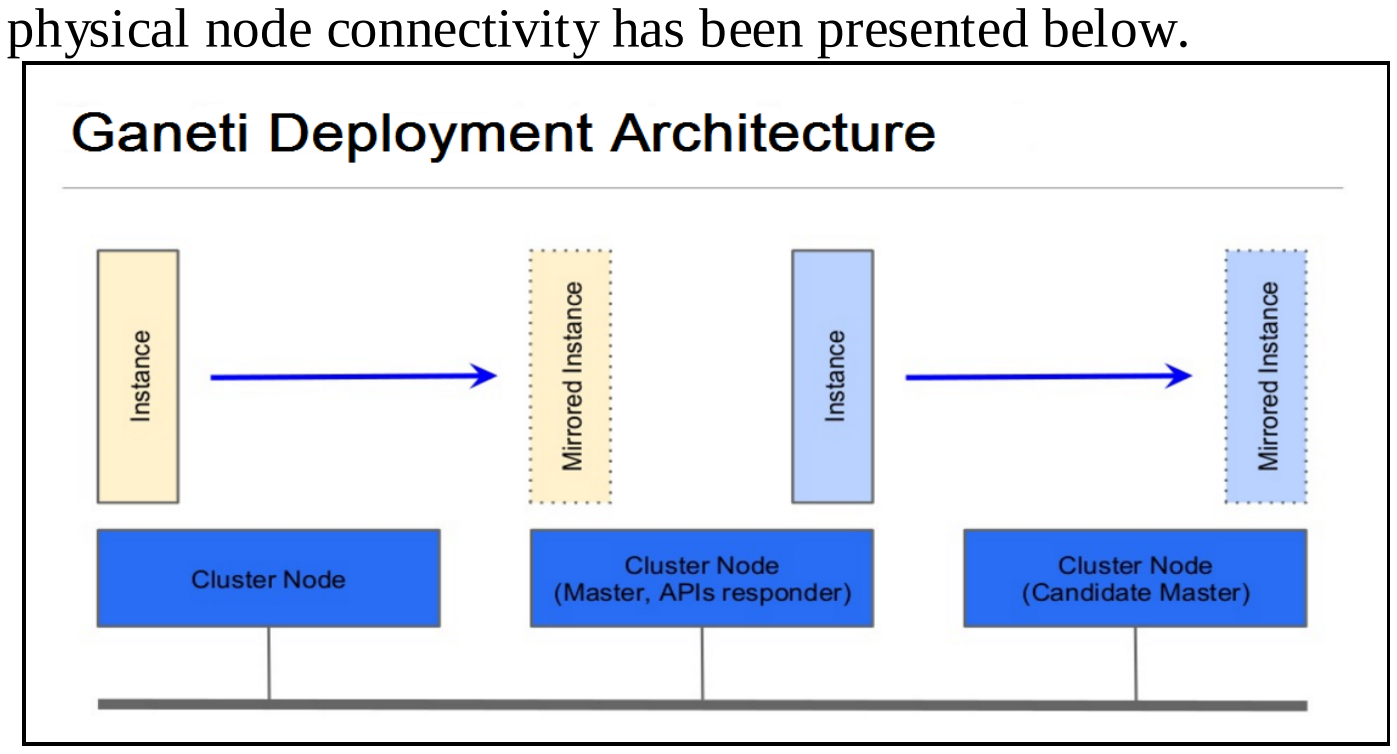

Figure2.1: Deployment Architecture

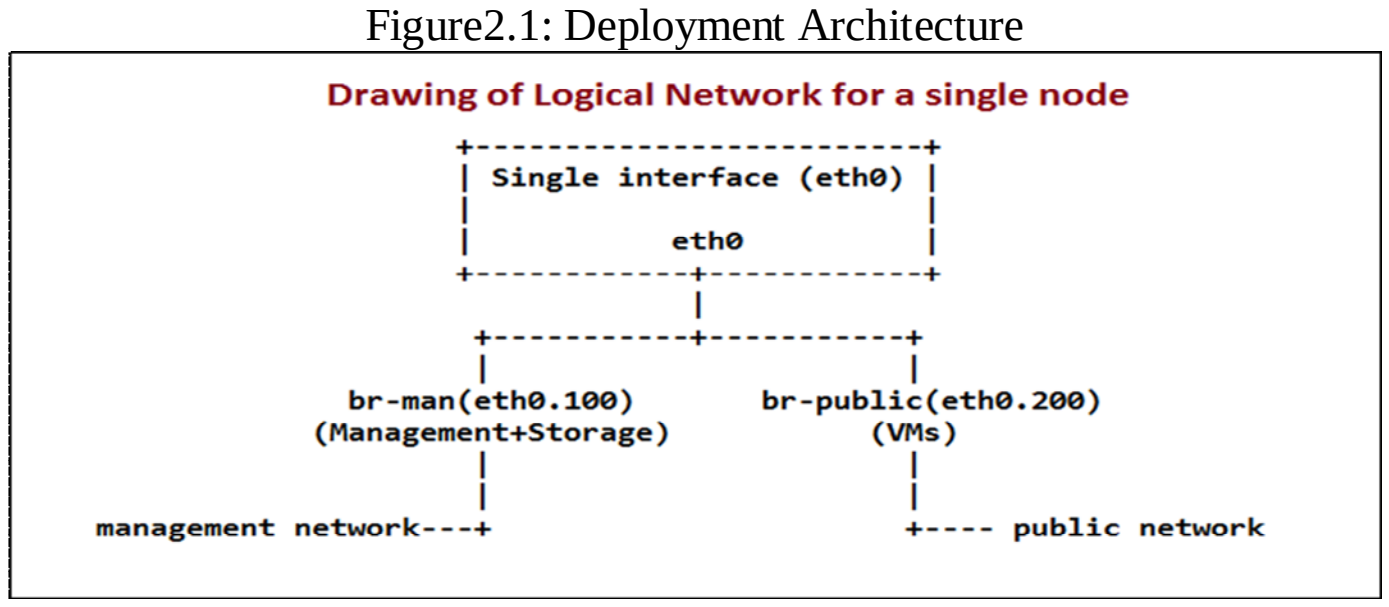

Figure 2.2: Network connectivity of a Physical Node

## 2.3 Installation of Base Operating System

This is mandatory for all nodes.

We have installed a clean, minimal operation system as standard OS. The only requirement we need to be aware of at this stage is to partition leaving enough space for a big (minimum 10GB) LVM volume group which will then host your instance file systems, if we want to use all Ganeti features. In this case we will install the base operating system on 10GB of our storage space and remaining storage space will leave un-partitioned for LVM use. The volume group name we use will be genetic.

### 2.4 Configure the Hostname

Look at the contents of the file /etc/hostname and check it contains the fully-qualified domain name, i.e. **node1.project.edu**

Now get the system to re-read this file:

```
# hostname -F /etc/hostname
```

Also check `/etc/hosts` to ensure that you have the both the fully-qualified name and the short name there, pointing to the correct IP address:

```
127.0.0.1    localhost
192.168.20.222    node1.project.edu    node1
```

### 2.5 Creating Logical Volume Manager

Type the following command:

```
# vgs
```

If it shows we have a volume group called 'ganeti' then skip to the next section, "Configure the Network". If the command is not found, then install the lvm2 package:

```
# apt-get install lvm2
```

Now, our host machine should have either a spare partition or a spare hard drive which we will use for LVM. If it's a second hard drive it will be `/dev/vdb` or `/dev/sdb`. Check which you have:

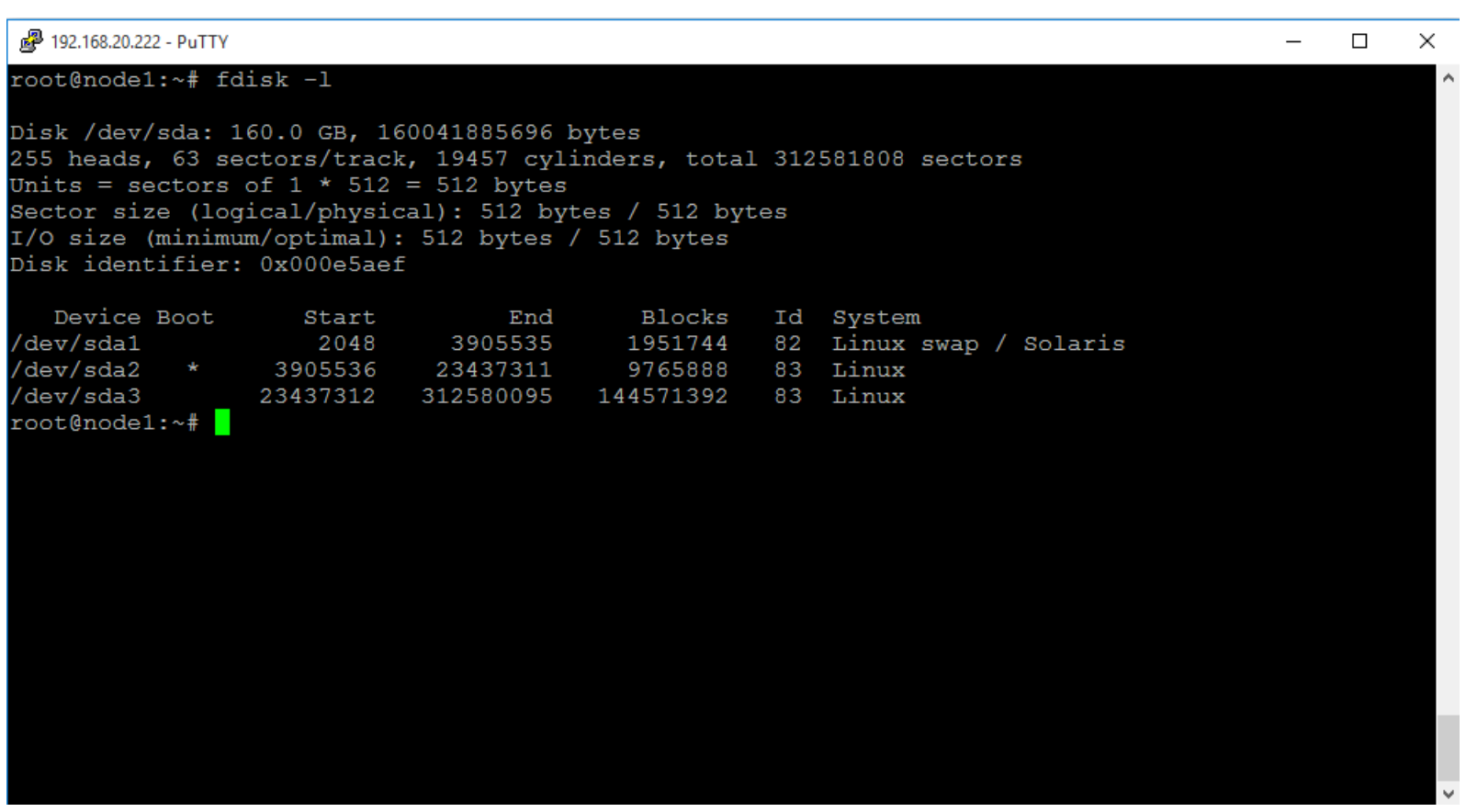

Figure 2.3: Checking available disks

Assuming `/dev/sda3` is spare; let's mark it as a physical volume for LVM:

```
# pvcreate /dev/sda3
# pvs   # should show the physical volume
```

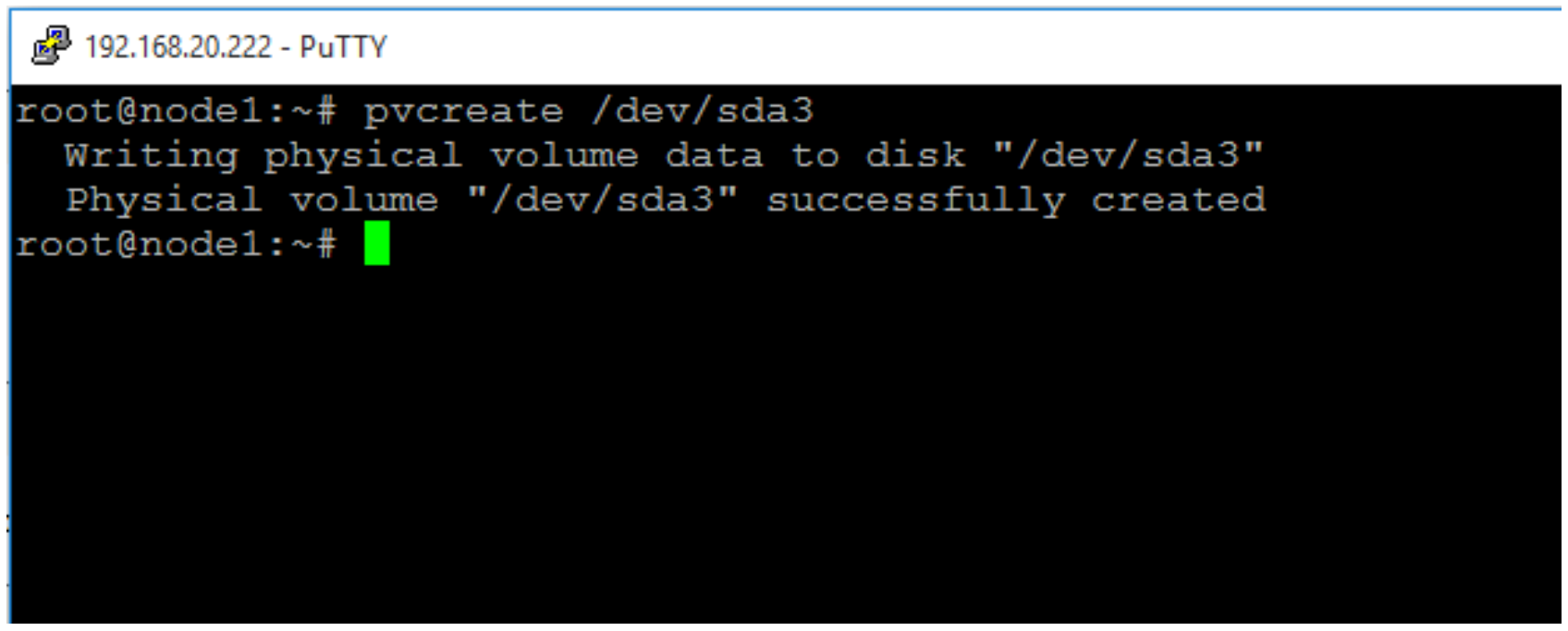

Figure 2.4: Physical Volume Create

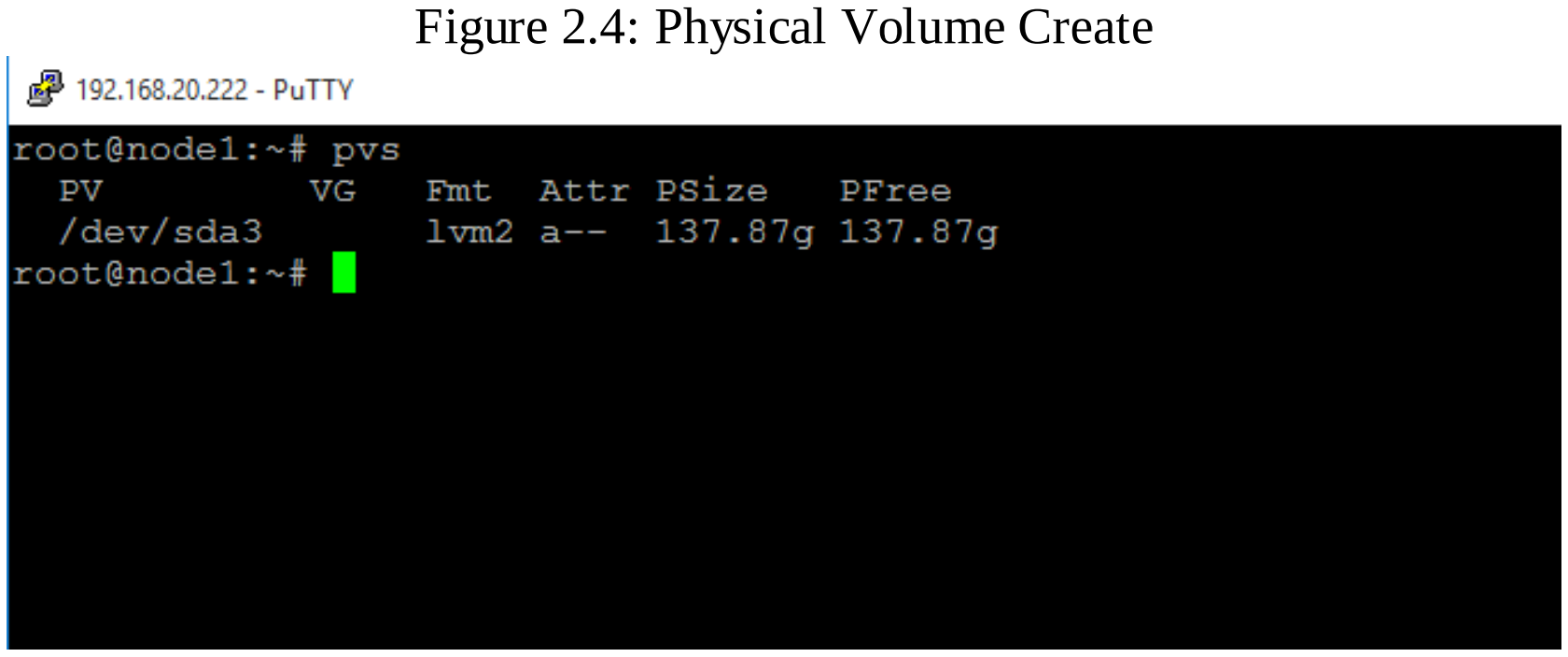

Figure 2.5: Physical volume check

Now we need to create a volume group called ganeti containing just this one physical volume. (Volume groups can be extended later by adding more physical volumes)

```
# vgcreate ganeti /dev/vdb
# vgs
```

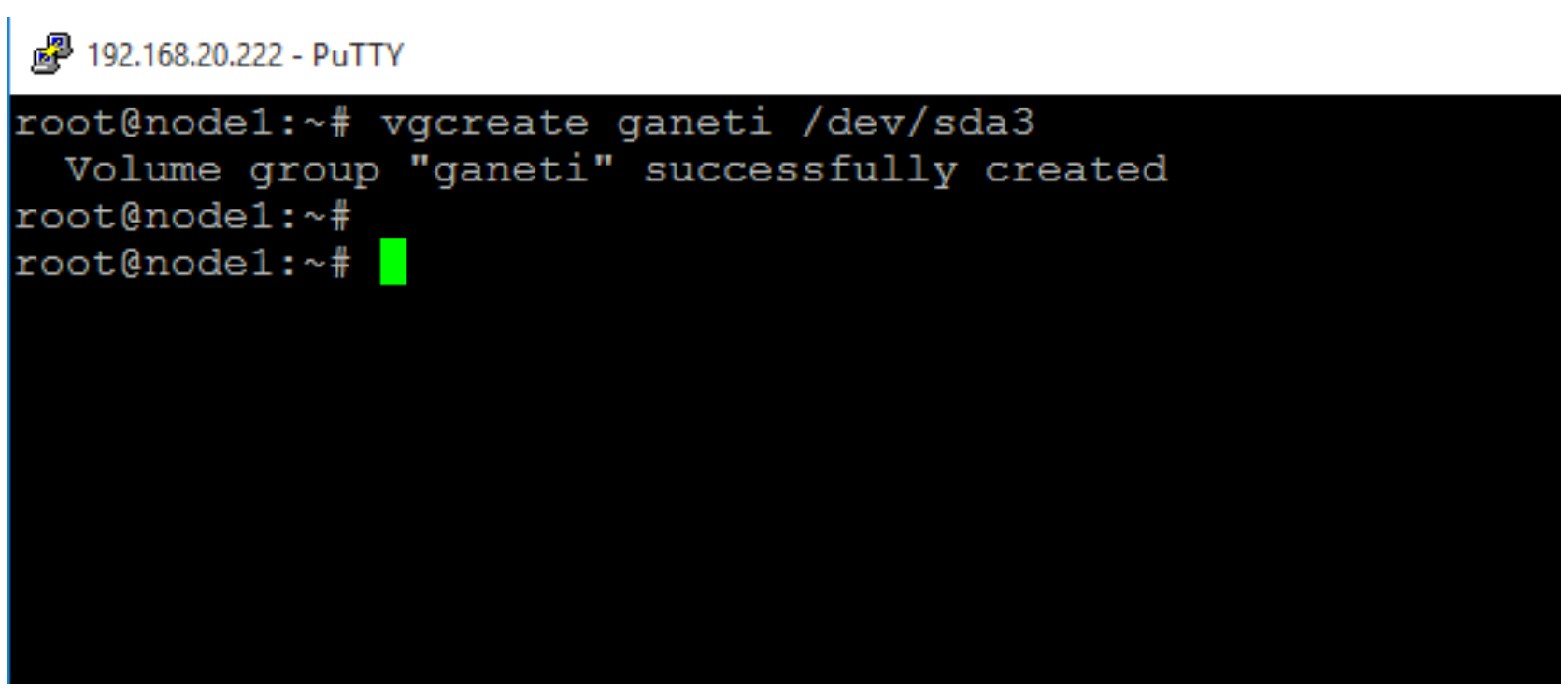

Figure 2.6: Volume Group Create

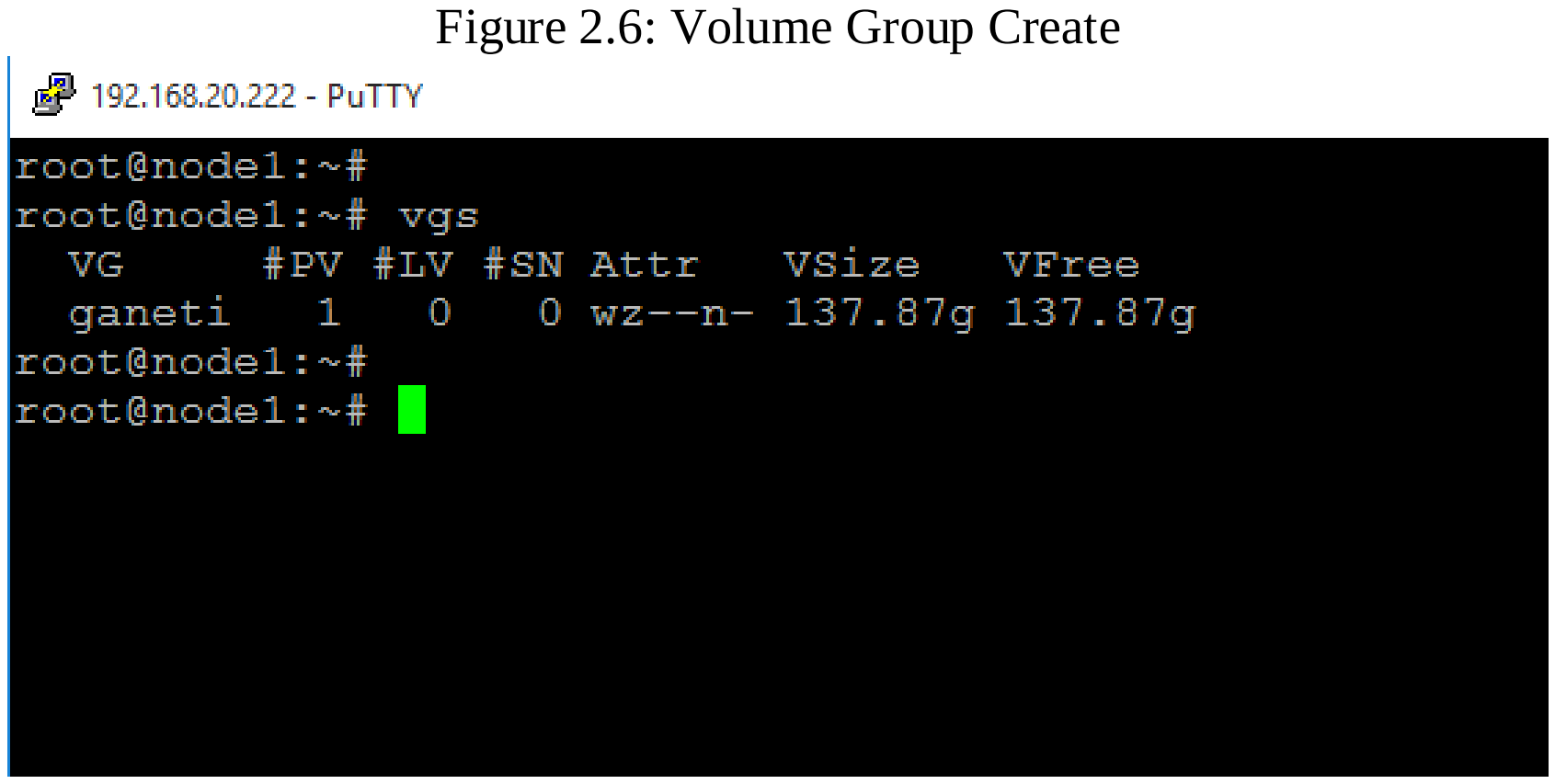

Figure 2.7: Volume Group Check

Note: on a production Ganeti server it is recommended to configure LVM not to scan DRBD devices for physical volumes. The document suggests editing `/etc/lvm/lvm.conf` and adding a reject expression to the filter variable, like this:

```
filter = [ "r|/dev/cdrom|", "r|/dev/drbd[0-9]+|" ]
```

## 2.6 Configure the Network

We're now going to reconfigure the network on our machine, so that we will be using VLANs. While it would be perfectly fine to use a single network for running virtual machines, there are a number of limitations, including: No separation between the networks used to manage the servers (management) and the disk replication network i.e. storage network. We will be using network-based disk replication. We'd like to keep the disk traffic separate from the management and service traffic Instead of using separate Ethernet cards, we'll use VLANs. In commodity hardware we usually have only one network interface.

We need to implement the networks: *management*, *replication*, and *service*.

Ideally, we would create two VLANs:

A management + Storage VLAN (vlan 100).

An external (or service) VLAN (vlan 200), where we will "connect" the virtual machines to publish them on internet.

*VLAN configuration*

To be on the safe side, let's install the vlan and bridge management tools (these should already have been installed by you earlier).

```
# apt-get install vlan bridge-utils
```

Let's make changes to the network configuration file for your system. If you remember, this is `/etc/network/interfaces`.

Edit this file, and look for the br-man definition for management and storage network and br-public for public VM network. This is the bridge interface you created earlier, and eth0 is attached to it. If should looks something like this:

```
# The loopback network interface
auto lo
iface lo inet loopback



auto br-man
iface br-man inet static
        address 192.168.20.222
        netmask 255.255.255.0
        gateway 192.168.20.1
        dns-namesevers 192.168.20.1
        bridge_ports eth0.100
        bridge_stp off
        bridge_fd 0
        post-up sys -w net.ipv6.conf.${IFACE}.disable_ipv6=1 #diabiling IPv6 for this interface


auto br-public
iface br-public inet manual
        bridge_ports eth0.200
        bridge_stp off
        bridge_fd 0
        post-up sys -w net.ipv6.conf.${IFACE}.disable_ipv6=1 #diabiling IPv6 for this interface
```

Figure 2.8: Network Interface configuration

## 2.7 Synchronize the clock

It's important that the nodes have synchronized time, so install the NTP daemon on every node:

```
# apt-get install ntp
```

## 2.8 Install the Ganeti software

Now install the software from the right package repository. How to do this depends on whether your machine is running Debian or Ubuntu. On Debian, the available version of ganeti is too old, but fortunately the current version is available in a back ports repository.

As root, create a file `/etc/apt/sources.list.d/wheezybackports.list` containing this one line: deb http://cdn.debian.net/debian/ wheezy-back ports main then refresh the index of available packages:

```
# apt-get update
```

Now, install the Ganeti software package. Note that the back ports packages are not used unless you ask for them explicitly.

```
# apt-get install ganeti/wheezy-back ports
```

This will install the current released version of Ganeti on our system; but any dependencies it pulls in will be the stable versions.

### 2.9 Setup DRBD

We'll now set up DRBD (Distributed Replicated Block Device), which will make it possible for VMs to have redundant storage across two physical machines. DRBD was already installed when we installed Ganeti, but we still need to change the configuration:

```
# echo "options drbd minor_count=128 usermode_helper=/bin/true"
>/etc/modprobe.d/drbd.conf
# echo "drbd" >>/etc/modules
# rmmod drbd  # ignore error if the module isn't already loaded
# modprobe drbd
The entry in /etc/modules ensures that drbd is loaded at boot
time.
```

### 2.10 Initialize the cluster - MASTER NODE ONLY

We are now ready to run the commands that will create the Ganeti cluster. Do this only on the MASTER node of the cluster.

```
# gnt-cluster init --master-netdev=br-man --enabled-
hypervisors=kvm -N link=br-public --vg-name ganeti
cluster.project.edu

# gnt-cluster modify -H
kvm:kernel_path=,initrd_path=,vnc_bind_address=0.0.0.0
```

Adding nodes to the cluster - MASTER NODE ONLY
So let's run the command to add the other nodes. Note the use of the -s option to indicate which IP address will be used for disk replication on the node we are adding.
Run this command only on the MASTER node of the cluster.

```
# gnt-node add node2.project.edu
```

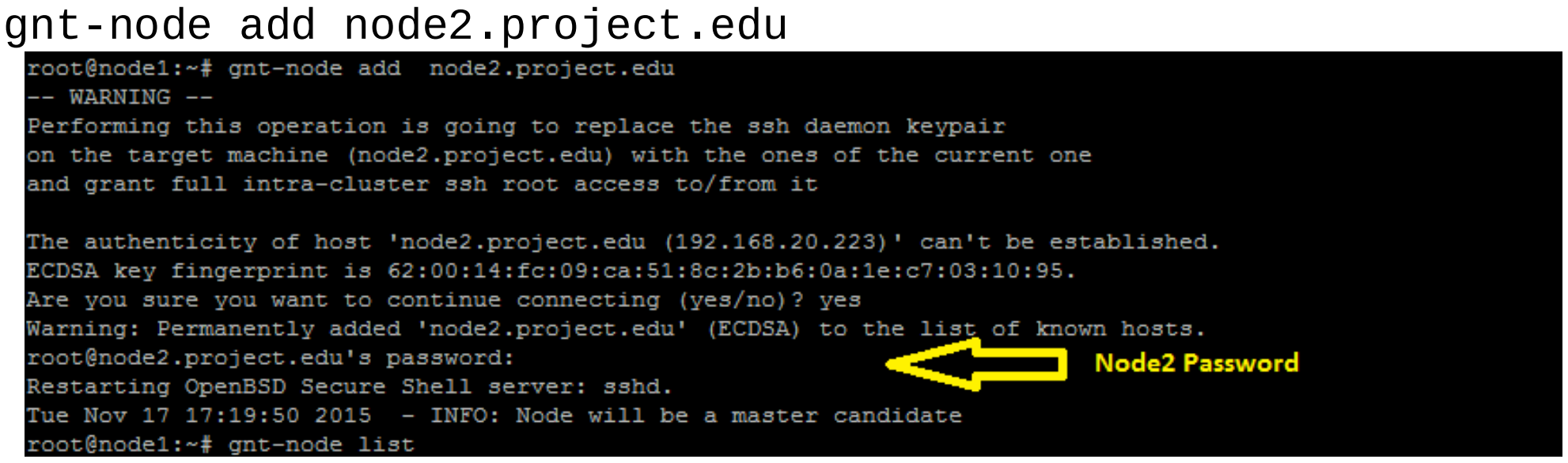

Figure 2.9: Add a node to the Cluster

We will be warned that the command will replace the SSH keys on the destination machine (the node you are adding) with new ones. This is normal.

```
                              -- WARNING --
Performing this operation is going to replace the ssh daemon keypair
on the target machine (hostY) with the ones of the current one and
grant full intra-cluster ssh root access to/from it
When asked if you want to continue connection, say yes:
The authenticity of host 'node2 (192.168.20.223)' can't be
established.
ECDSA key fingerprint is
a1:af:e8:20:ad:77:6f:96:4a:19:56:41:68:40:2f:06.
Are you sure you want to continue connecting (yes/no)? yes
When prompted for the root password for node2, enter it:
Warning: Permanently added 'node2' (ECDSA) to the list of known hosts.
root@node1's password:
```

You may see the following informational message; you can ignore it:
Restarting OpenBSD Secure Shell server: sshd.
Rather than invoking init scripts through `/etc/init.d`, use the service utility, e.g. `service ssh restart`
Since the script you are attempting to invoke has been converted to an Upstart job, you may also use the stop and then start utilities,
e.g. stop ssh ; start ssh. The restart utility is also available.

```
ssh stop/waiting
ssh start/running, process 2921
```

The last message you should see is this:
Tue Nov 17 17:19:50 2015 - INFO: Node will be a master candidate
This means that the machine you have just added into the node (hostY) can take over the role of configuration master for the cluster, should the master (hostX) crash or be unavailable.
Check the node has been added in cluster or not by following command:

```
#gnt-node list
```

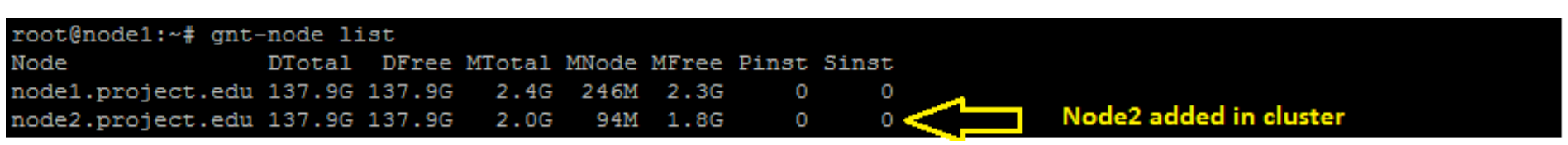

```
root@node1:~# gnt-node list
Node              DTotal DFree  MTotal MNode MFree Pinst Sinst
node1.project.edu 137.9G 137.9G   2.4G  246M  2.3G     0     0
node2.project.edu 137.9G 137.9G   2.0G   94M  1.8G     0     0
```

Figure 2.10: Node list

Now add the remaining node in our cluster and check the status again.

## 2.11 Installing OS definition - ALL NODES

We need to install a support package called ganeti-instance-image. This provides ganeti with an "OS definition" - a collection of scripts which ganeti uses to create, export and import an operating system.

The package can be installed as follows: do this on all nodes in your cluster.

```
# wget https://code.osuosl.org/attachments/download/2169/ganeti-instance-image_0.5.1-1_all.deb
# dpkg -i ganeti-instance-image_0.5.1-1_all.deb
```

## 2.12 Update the OS definition - MASTER ONLY

First wait until the other (slave) nodes in our cluster have installed the ganeti-instance-image package. Instance-image needs to be told how to install or re-install the operating system. It can be configured to do this by unpacking an image of an already-installed system (in tar, dump or qcow2 format), but in our case we just want to do a manual install from a CD image.

On the master node, as root create a file `/etc/ganeti/instance-image/variants/cd.conf` with the following contents:

```
CDINSTALL="yes"
NOMOUNT="yes"
```

Aside: the full set of settings you could put in this file are listed in `/etc/default/ganeti-instance-image`, but don't edit them there

Now edit `/etc/ganeti/instance-image/variants.list` so it looks like this:

```
default
cd
```

Copy these two files to the other nodes:

```
# gnt-cluster copyfile /etc/ganeti/instance-image/variants/cd.conf
# gnt-cluster copyfile /etc/ganeti/instance-image/variants.list
```

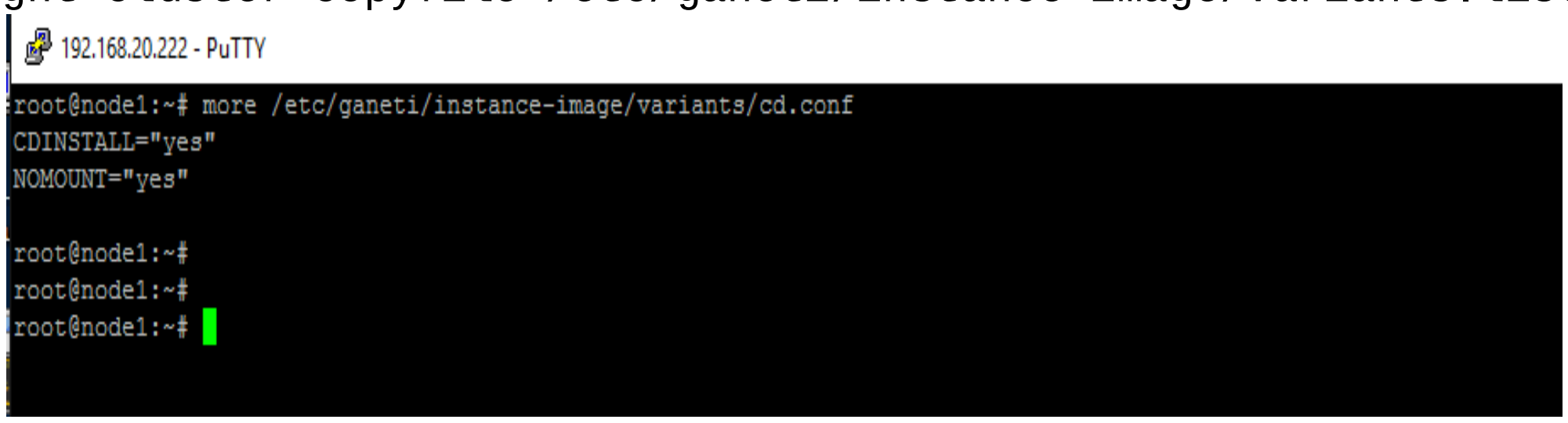

Figure 2.11: Variants Configuration

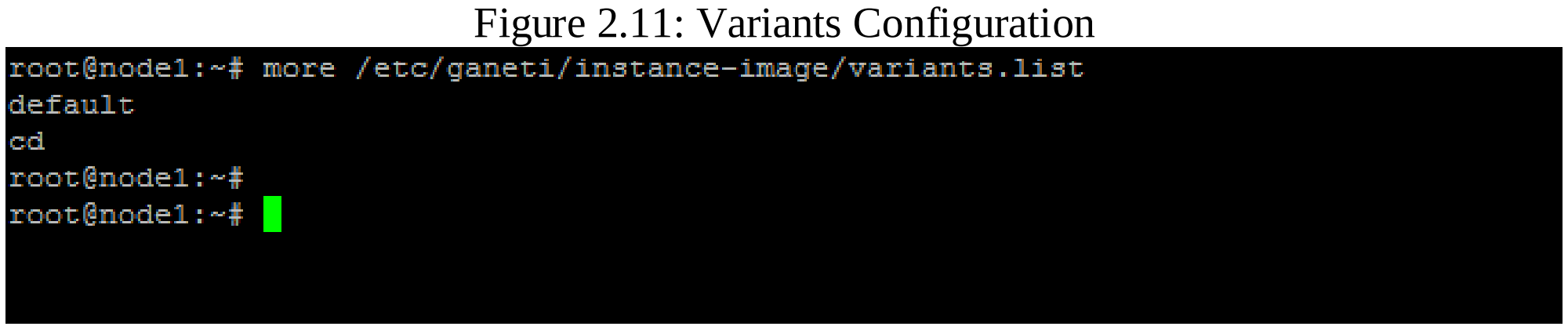

Figure 2.12: Variants check

Still on the master, check that the "image+cd" variant is available.

```
# gnt-os list
Name
debootstrap+default
image+cd                << THIS ONE
image+default
```

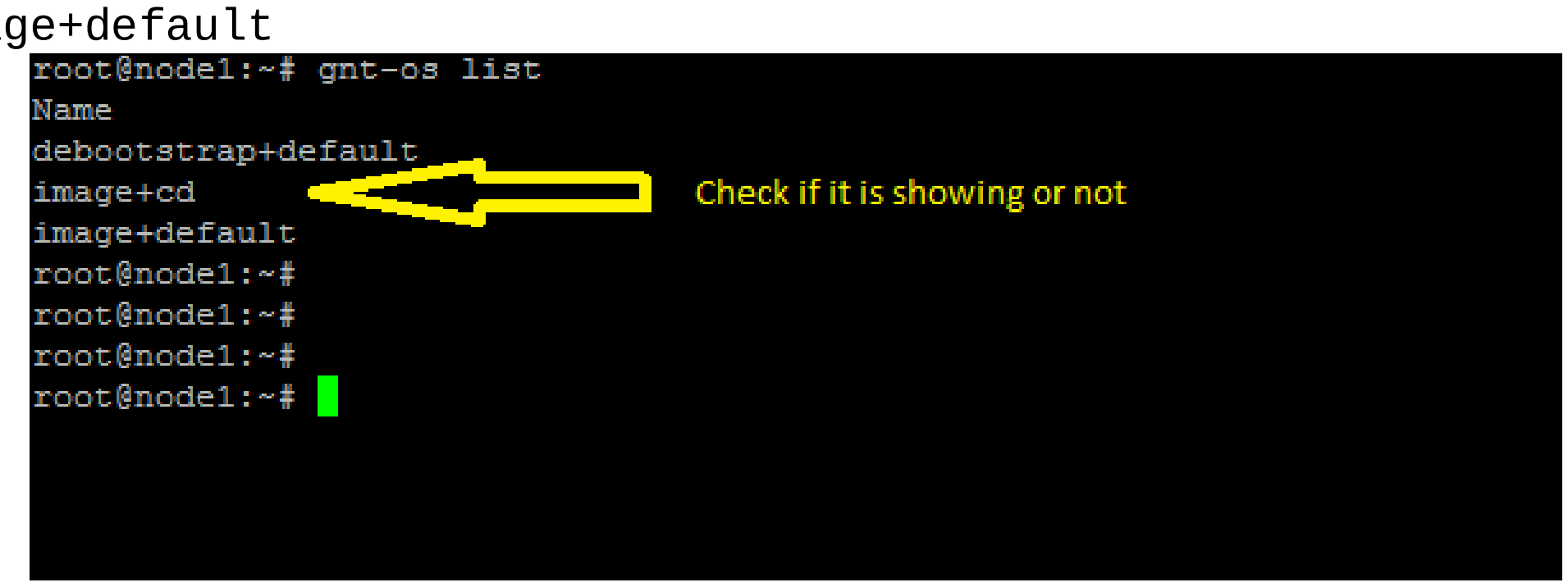

Figure 3.13: Boot source for instances

### 2.13 Distributing ISO images - ALL NODES

If using DRBD, the ISO images used for CD installs must be present on all nodes in the cluster, in the same path. You could copy them to local storage on the master node, and then use gnt-cluster copy file to distribute them to local storage on the other nodes. However to make things simpler, we've made all the ISO images available on an NFS share (Network File Service), which you can attach. On every node, create a empty directory `/iso`:

```
# mkdir /iso
```

Now copy a test OS iso in /iso directory. We have copied a debian iso image for test. Now send the iso image to every node by following command:

```
# gnt-cluster copyfile  /iso/debian-7.9.0-amd64-netinst.iso
```

### 2.14 Creation of instance - EVERYONE ON MASTER NODE

For example, if you working on host3 then you will have to login to host1 (your cluster's master node). You will then create a VM called testvm.project.edu and instruct ganeti to create it on your host using the flag -n node3.project.edu. To create new an instance, run the following command. (Note that we don't start it yet, because we want to temporarily attach the CD-ROM image at start time).

```
# gnt-instance add -t drbd -o image+cd -s 4G -B
minmem=256M,maxmem=512M --no-start --no-name-check --no-ip-check
testvm.project.edu
```

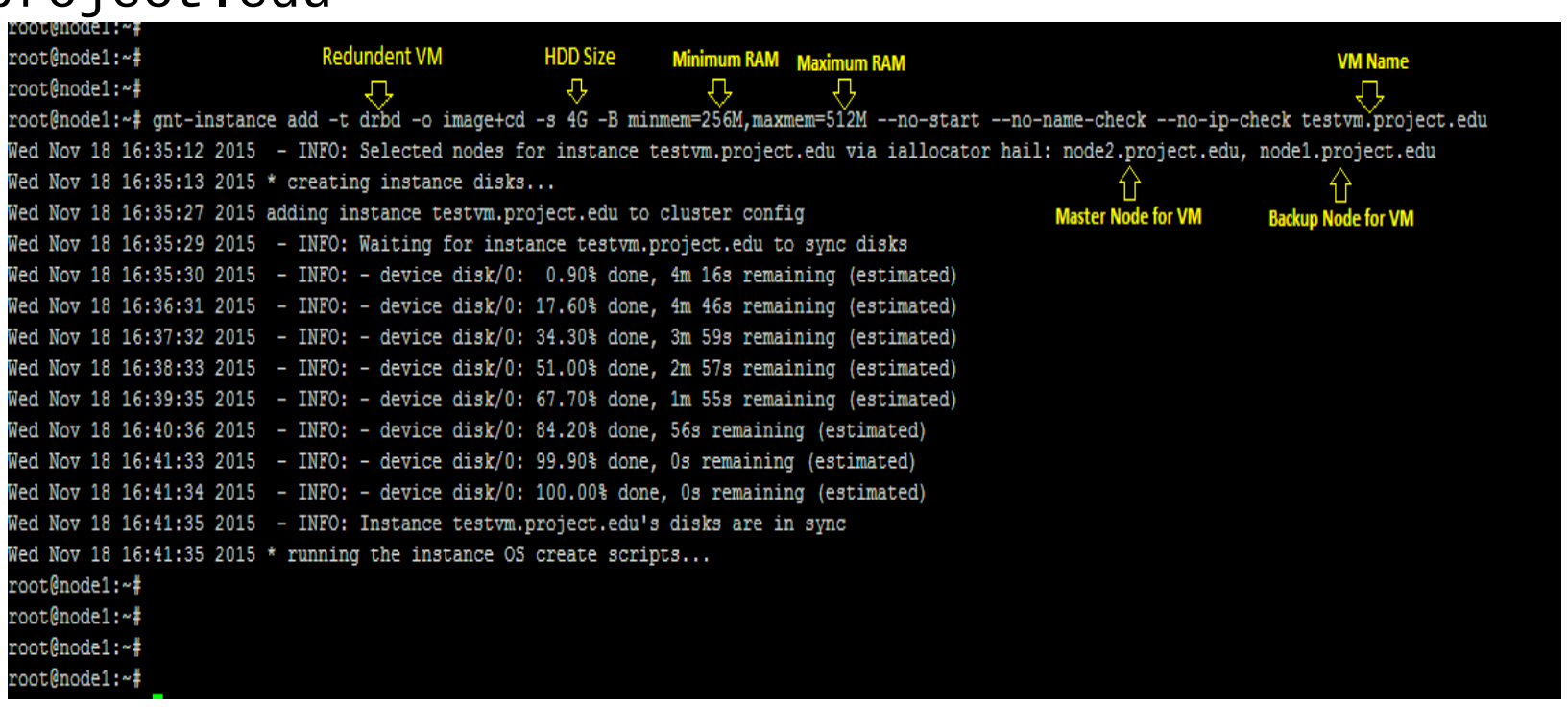

Figure 2.13: Instance create

**Explanation:**

-t drbd means replicated LVM (replicated with DRBD)

-o image+cd means to use the OS definition ganeti-instance-image, with the cd variant we created

-s 4G means to create a 4GB disk drive

-B minmem=256M,maxmem=512M sets the memory limits for this VM. It will try to run it with 512M, but if not enough memory is available it may shrink it down to 256MB.

--no-start means don't start the VM after creating it

--no-name-check means don't check that testvm.project.edu exists in the DNS (because it doesn't!)

--no-ip-check means if you found the name in the DNS, don't check that the IP address is not in use the final parameter is the name of the instance. It would be good practice to use a fully-qualified domain name for this.

You will see some messages about creating the instance being created.

## 2. RESULTS

### 3.1 Run an instance

Now start the VM using the following command, which attaches the CD-ROM temporarily and boots from it:

```
# gnt-instance start -H
boot_order=cdrom,cdrom_image_path=/iso/debian-7.9.0-amd64-
netinst.iso testvm.project.edu
```

Waiting for job 332 for testvm.project.edu ...

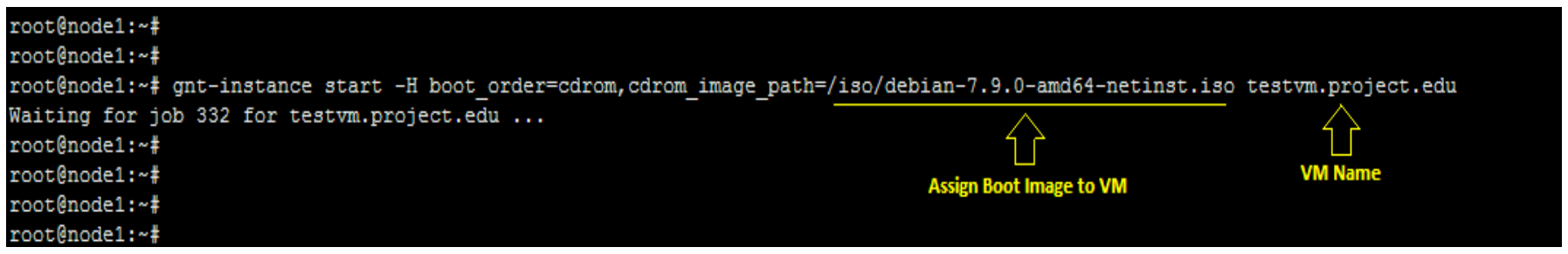

Figure 3.1: Run an instance

### 3.2 Verify the configuration of your cluster

Again only on the MASTER node of the cluster:

```
# gnt-cluster verify
```

This will tell you if there are any errors in your configuration.

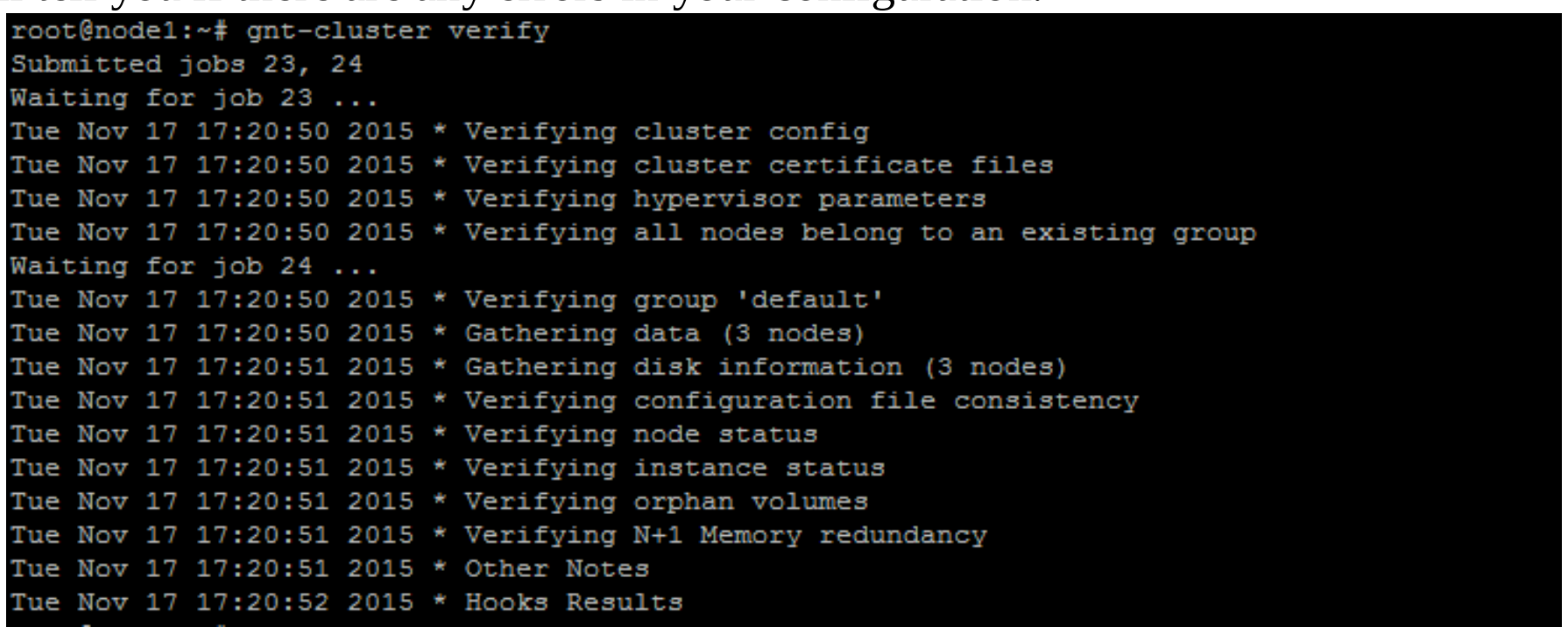

Figure 3.2: Verify Cluster

### 3.3 Check detail information about an instance

Ganeti will assign a port for console access for the created VM so that we can install the operating system on it remotely. Here is how to check it.

```
# gnt-instance info testvm.project.edu
```

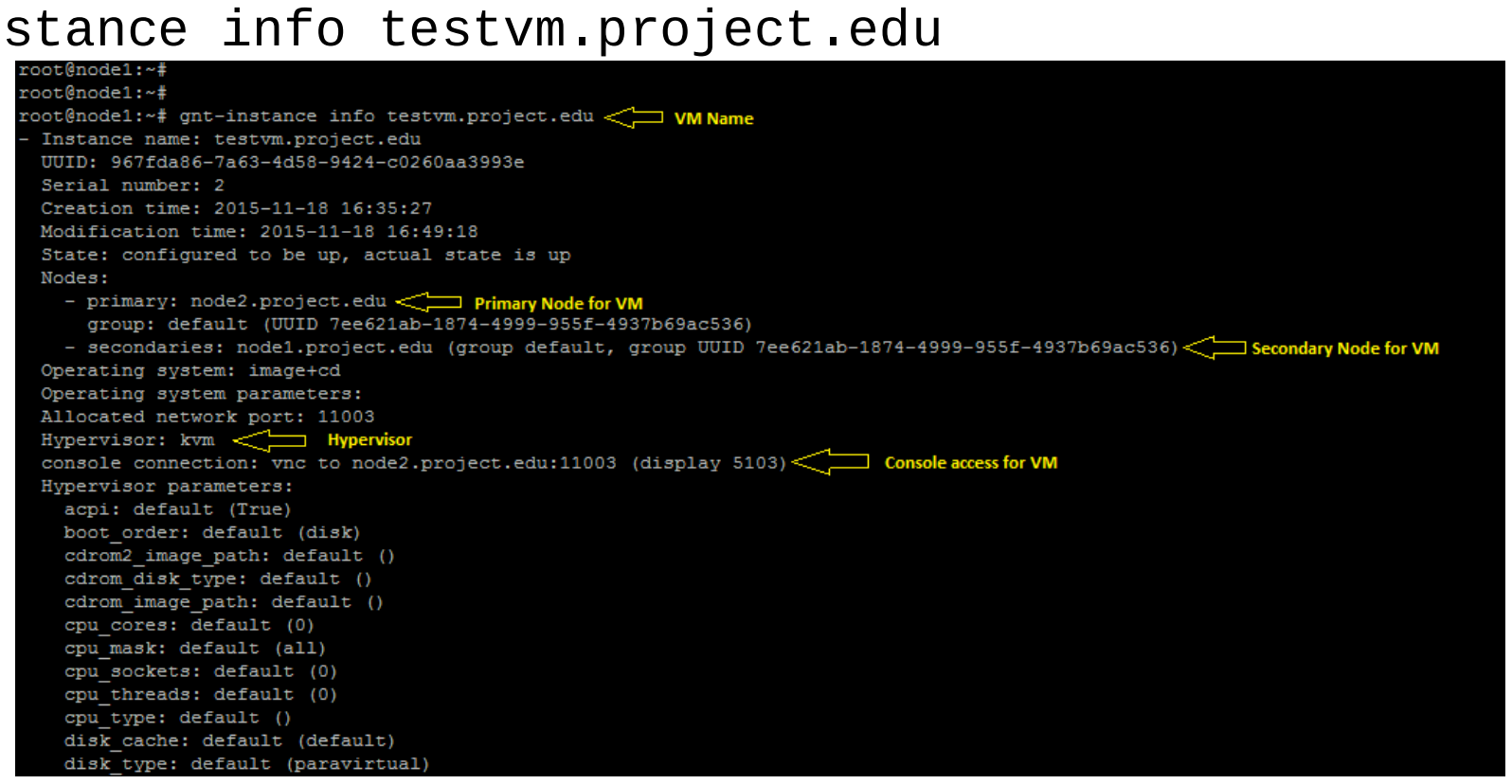

Figure 3.3: Information about an instance-1

More information about testvm.project.edu

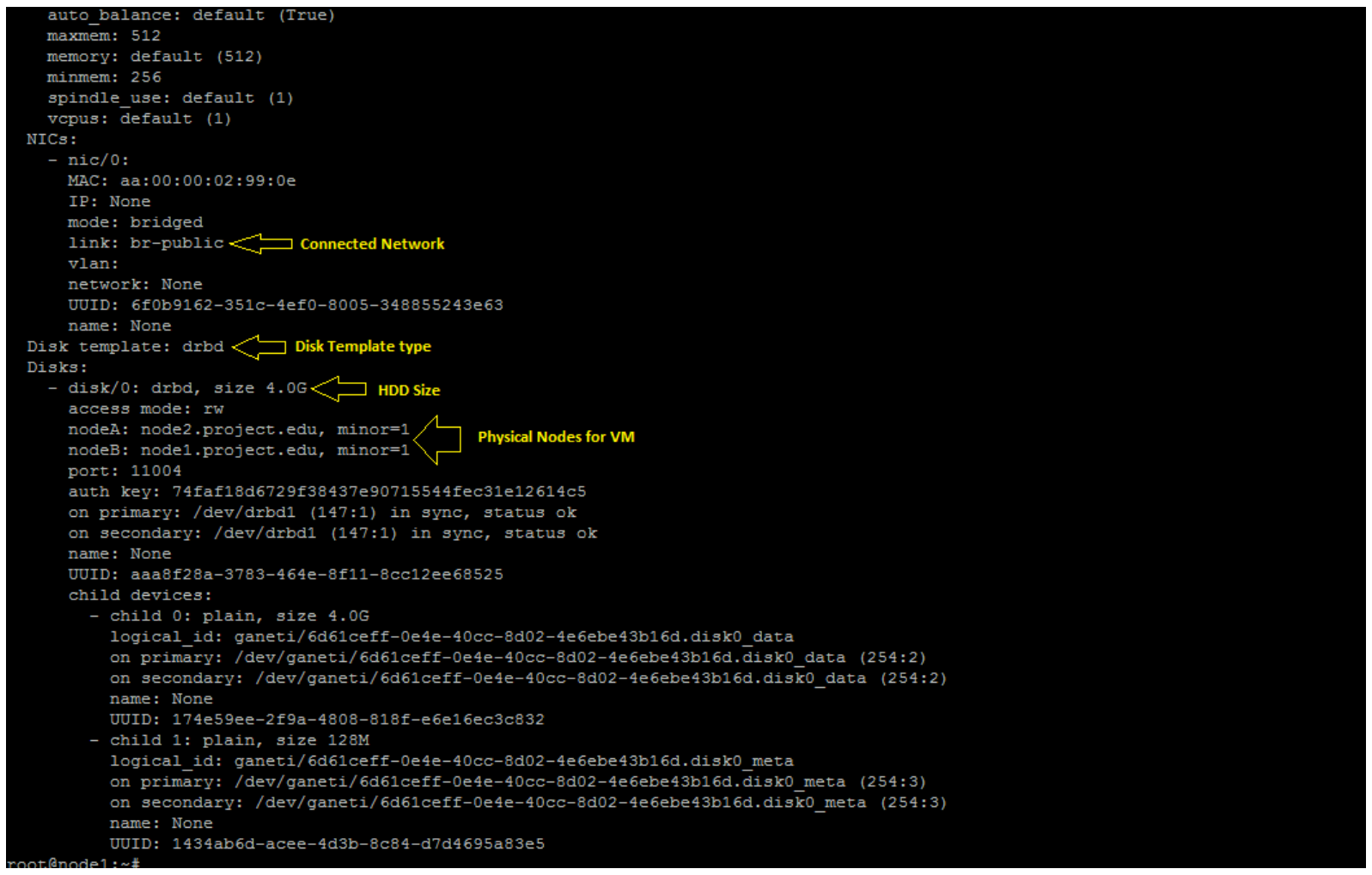

Figure 3.4: Information about an instance-2

## 3.4 Install a guest Operating System in an instance

We can see the console access for the VM is node2.project.edu:11003. We will use a VNC viewer to access the VM and install the operating system on it.

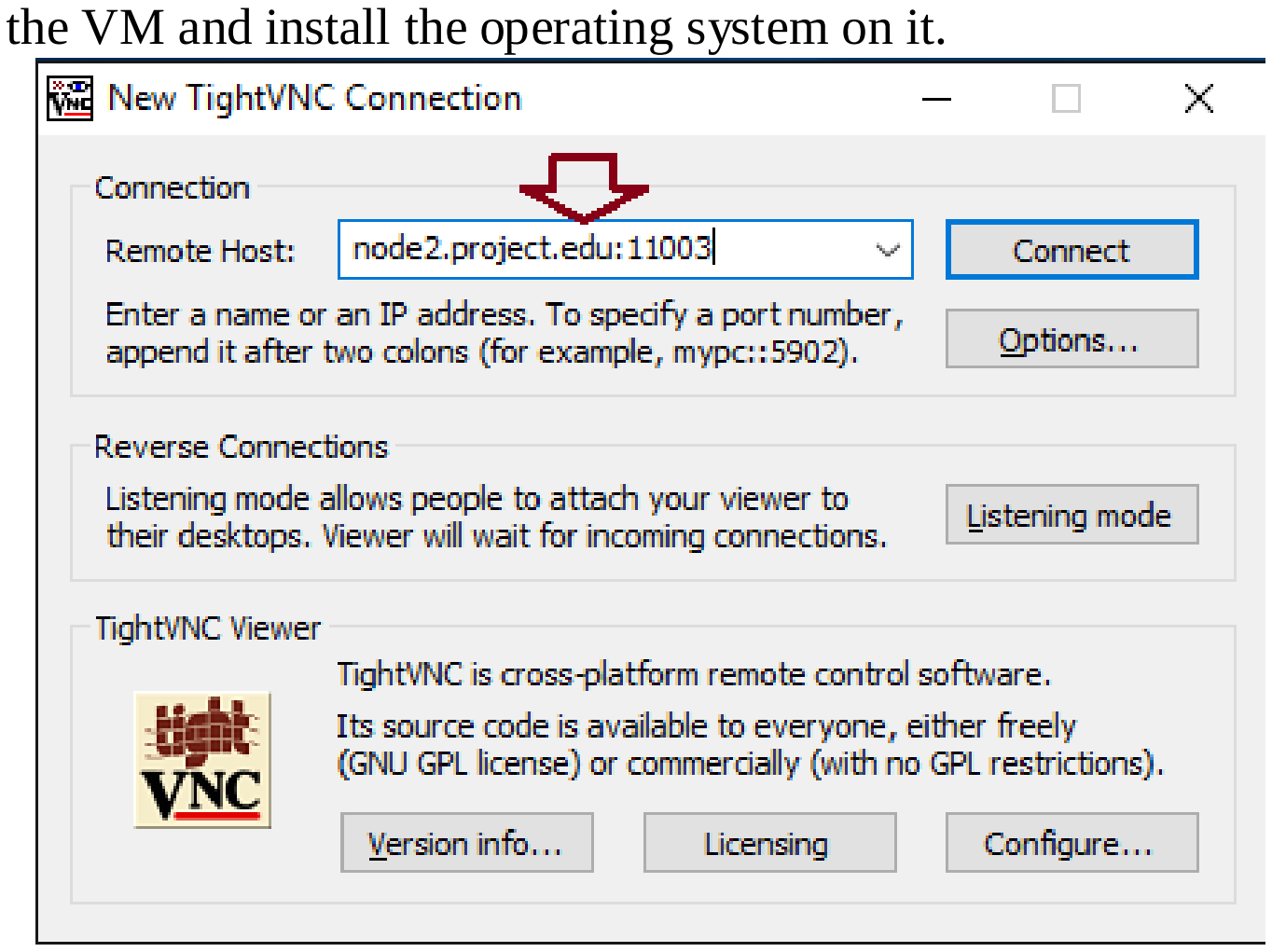

Figure 3.5: Connect an instance by VNC Viewer

By clicking "Connect" button the console will appear in front of us and we will install the OS on `testvm.project.edu` instance with the IP address of `192.168.20.232`.

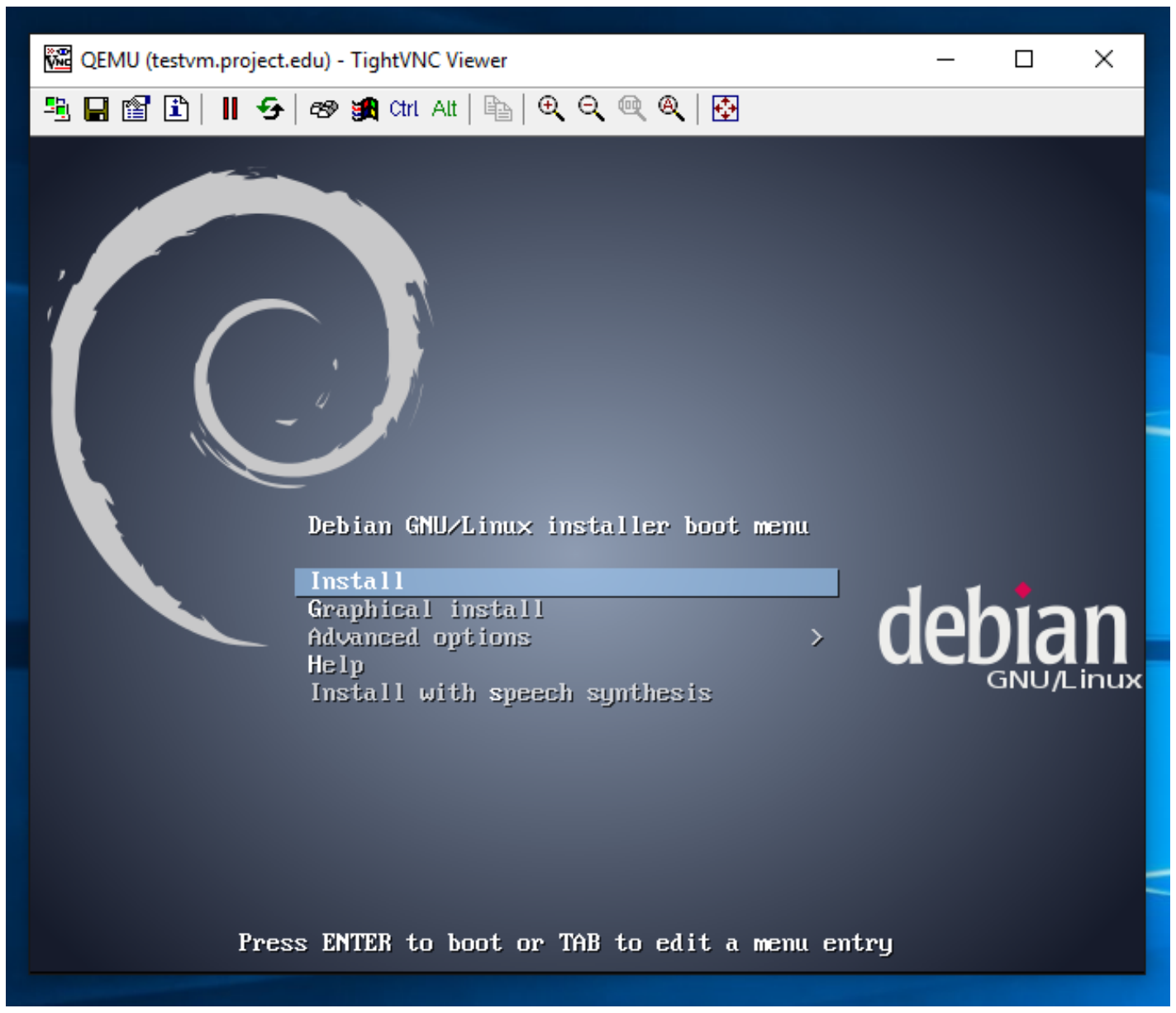

Figure 3.6: Install Guest Operating System in an Instance

### 3.5 Changing network of an Instance

We may not need to run this, but if we want to we can. Let's say we have informed our cluster `“br-public”` as the default network for every instance. Now we are connected to the "br-man" network. As a result we cannot access the VM from remote network as "br-man" is not published to internet.

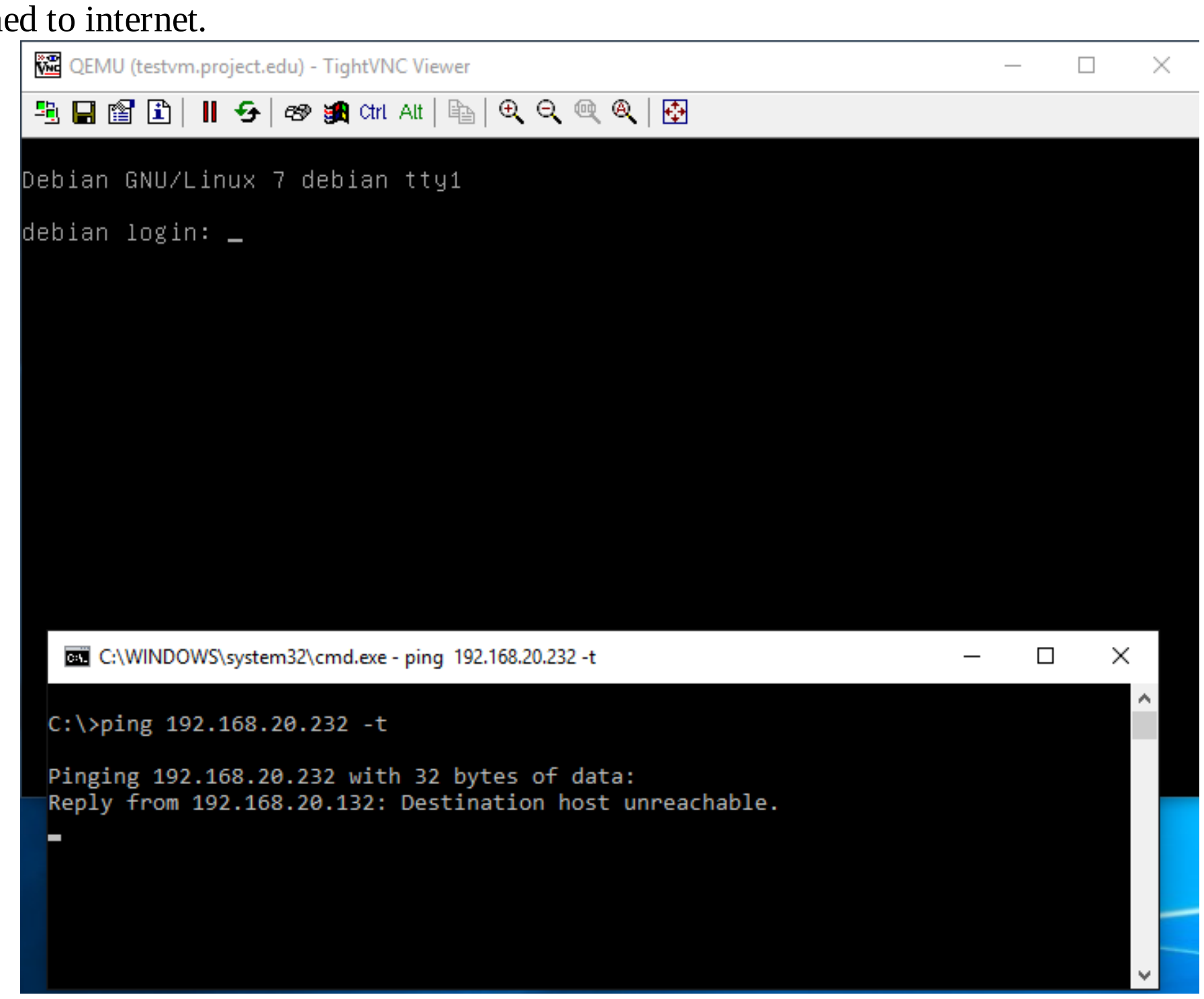

Figure 3.7: Connectivity check for an instance

Now we need to change the network of the instance to `“br-man”` from `“br-public”`. Here is how to do that:

Moving the network interface 0 to another network:

```
# gnt-instance modify --net 0:modify,link=br-man --hotplug testvm.project.edu
```

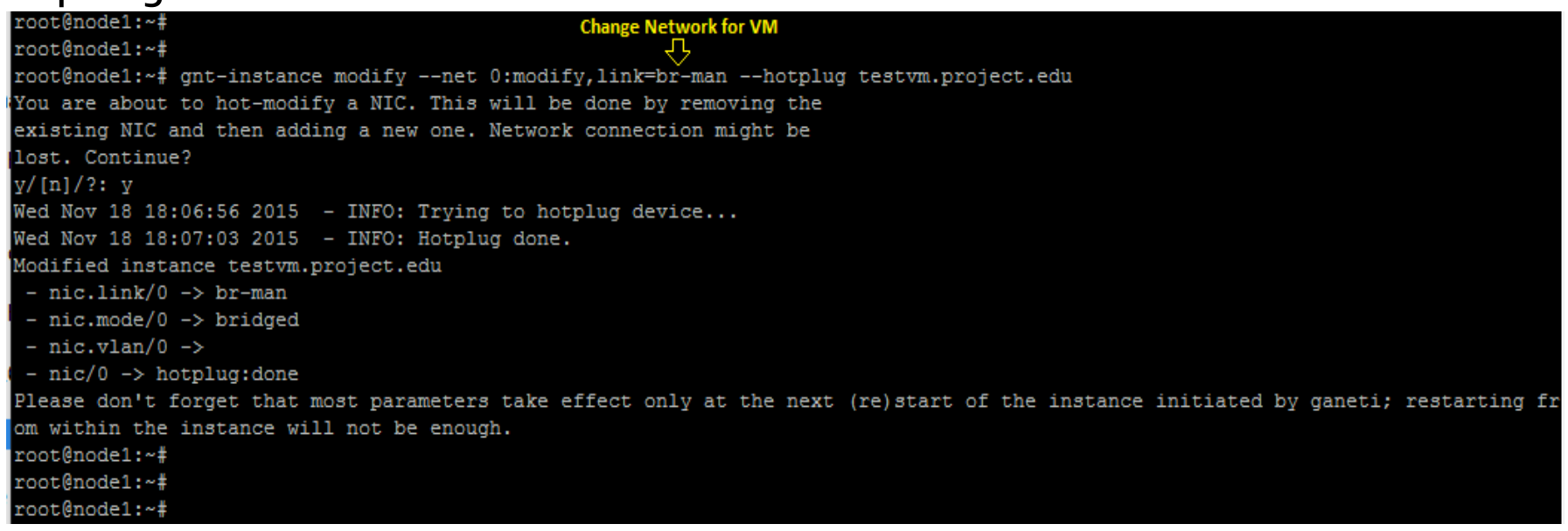

Figure 3.8: Change the network for an instance

Try to do this to move the network interface of one of the instances you created earlier, onto the br-man. After successfully shifting the network, now we can access the instance without any problem.

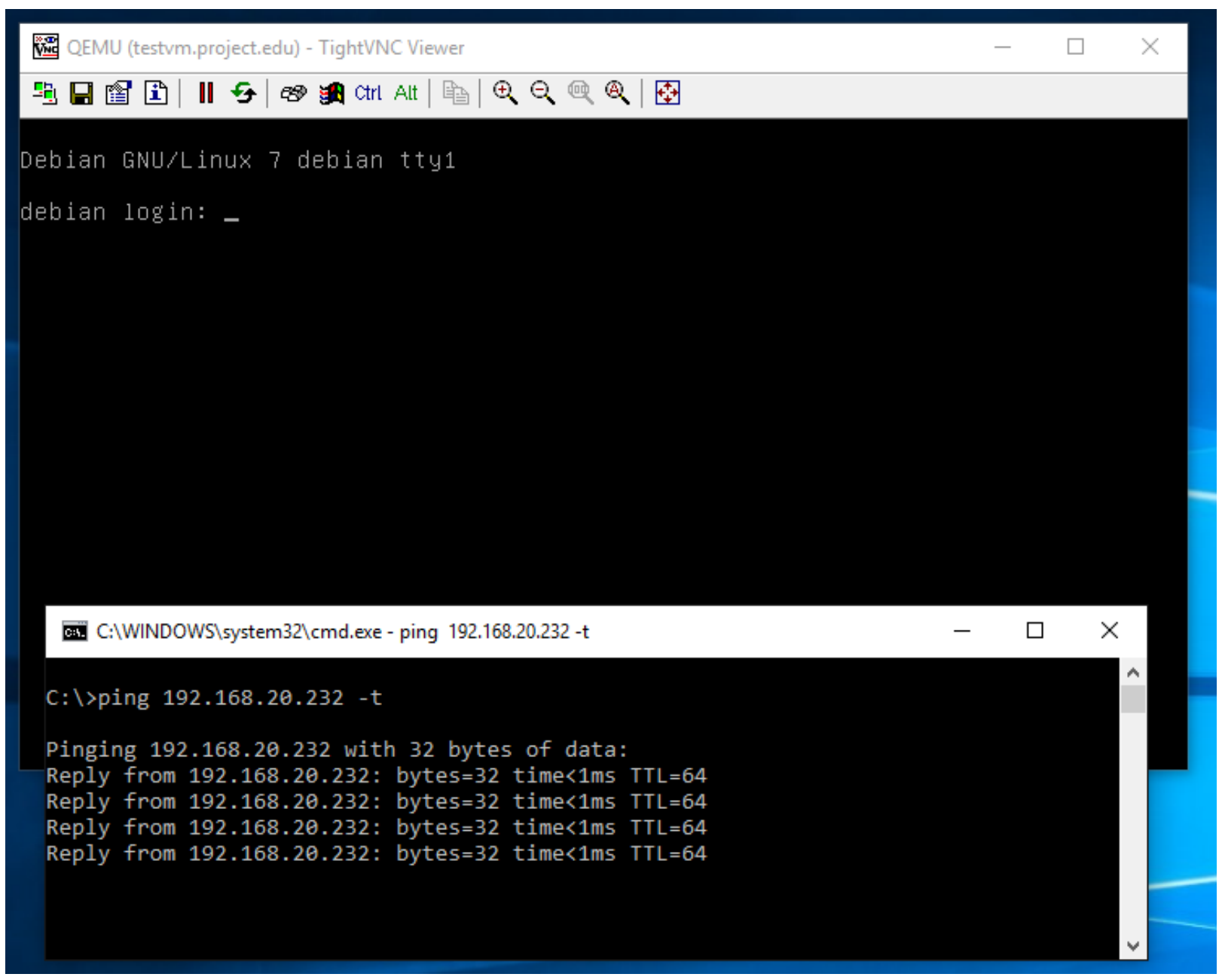

Figure 3.9: Check availability after changing the network of an instance

### 3.6 Testing Service Migration between two instances

In some cases we may need to change the service of an instance to backup node. Let's say we have an instance running with DRBD replication on `node2.project.edu` as primary node and `node1.project.edu` as backup node. Now we have a situation and we need to shut down `node2.project.edu` for maintenance. We should do it without interrupting the users from their service. So we will be migrated the service to `node1.project.edu` and shut down `node2.project.edu` for maintenance.

```
# gnt-instance migrate testvm.project.edu
```

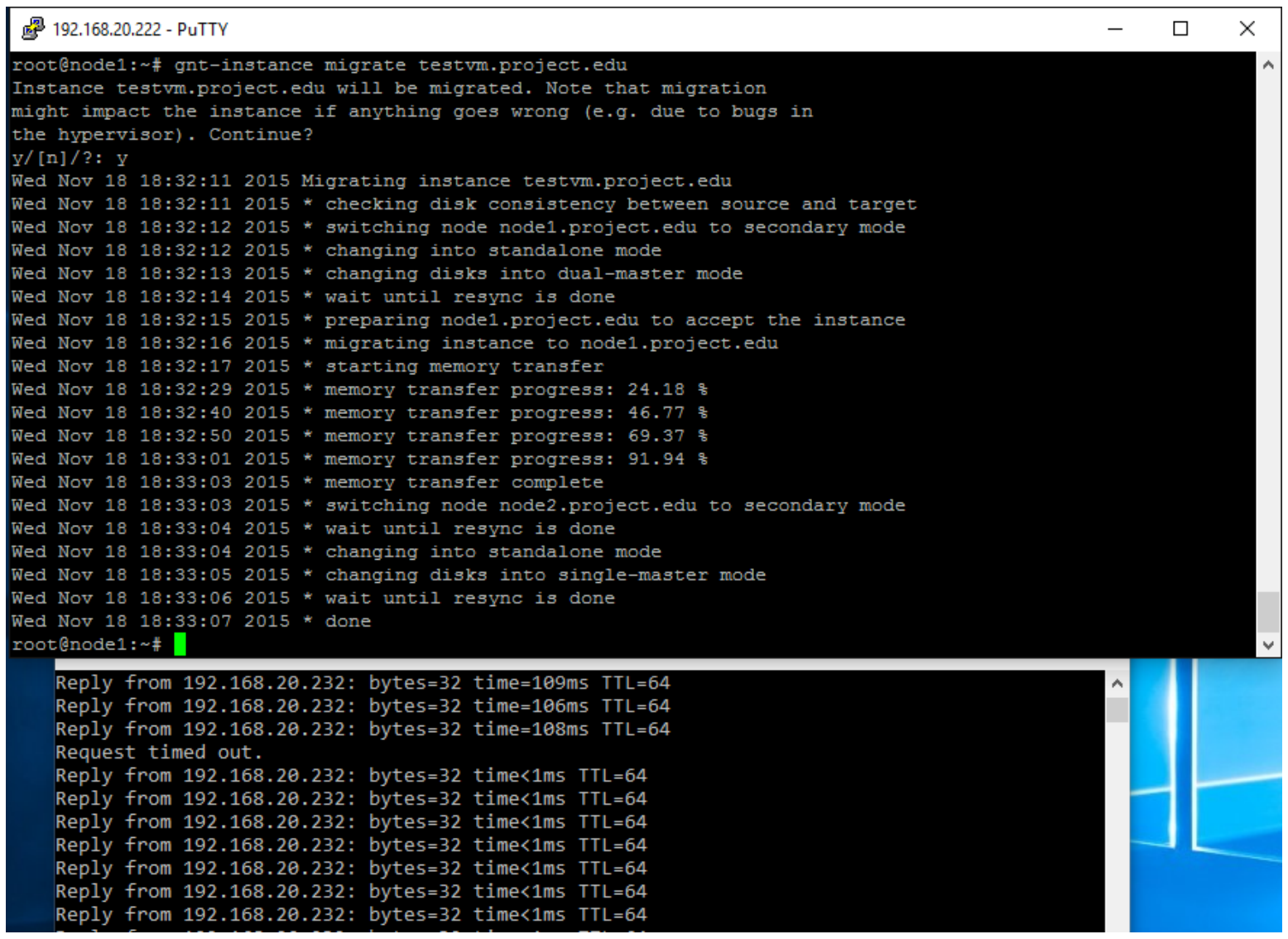

Figure 3.10: Live service migration of an instance

### 3.7 Instance Failover Scenario

Suppose we have an instance running on `node2.project.edu` as its primary node and `node1.project.edu` as its backup node. Suddenly disaster happens; `node2.project.edu` has failed and went down.

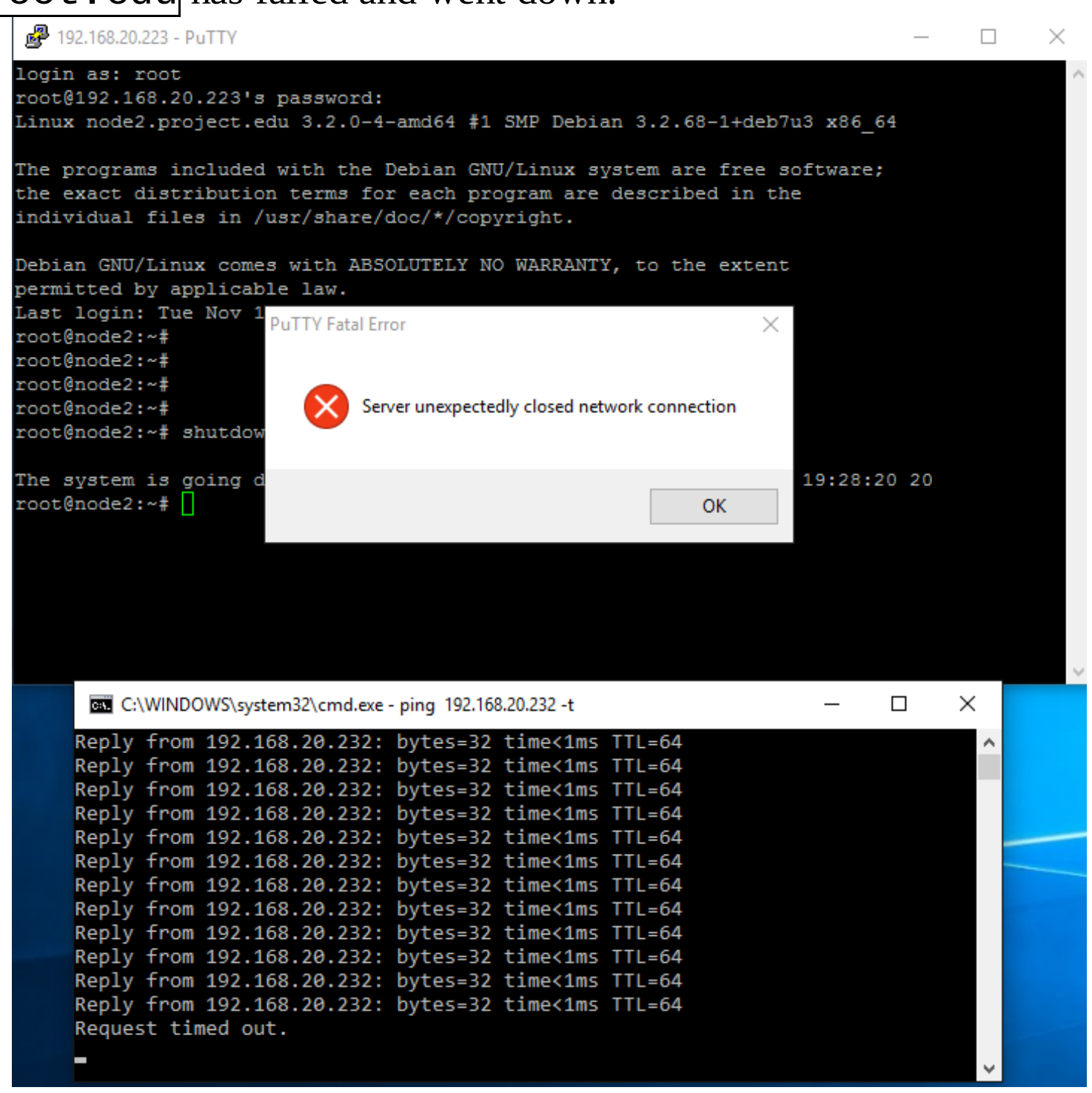

Figure 3.11: Failover of an instance

Instances' running on that node has been down and services stopped. But we can make the service alive within short time without losing any data of that instance by following command.

```
# gnt-instance failover –ignore-consistency testvm.project.edu
```

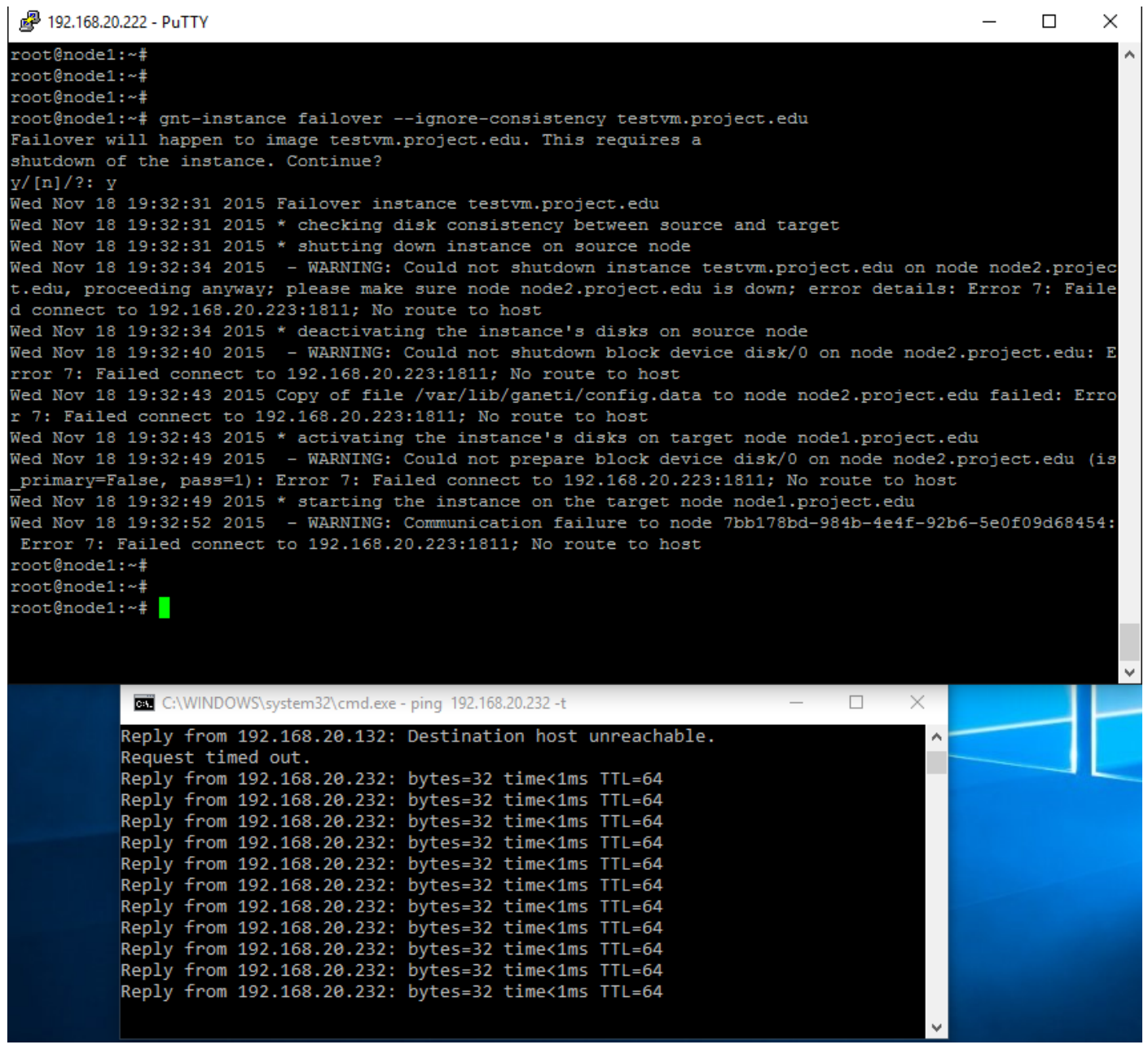

Figure 3.12: Recovery of an instance after failure

Now if we check the instance list we can see the instance is running on `node1.project.edu` as its primary node.

```
# gnt-instance list -o name,pnode,snodes,status
```

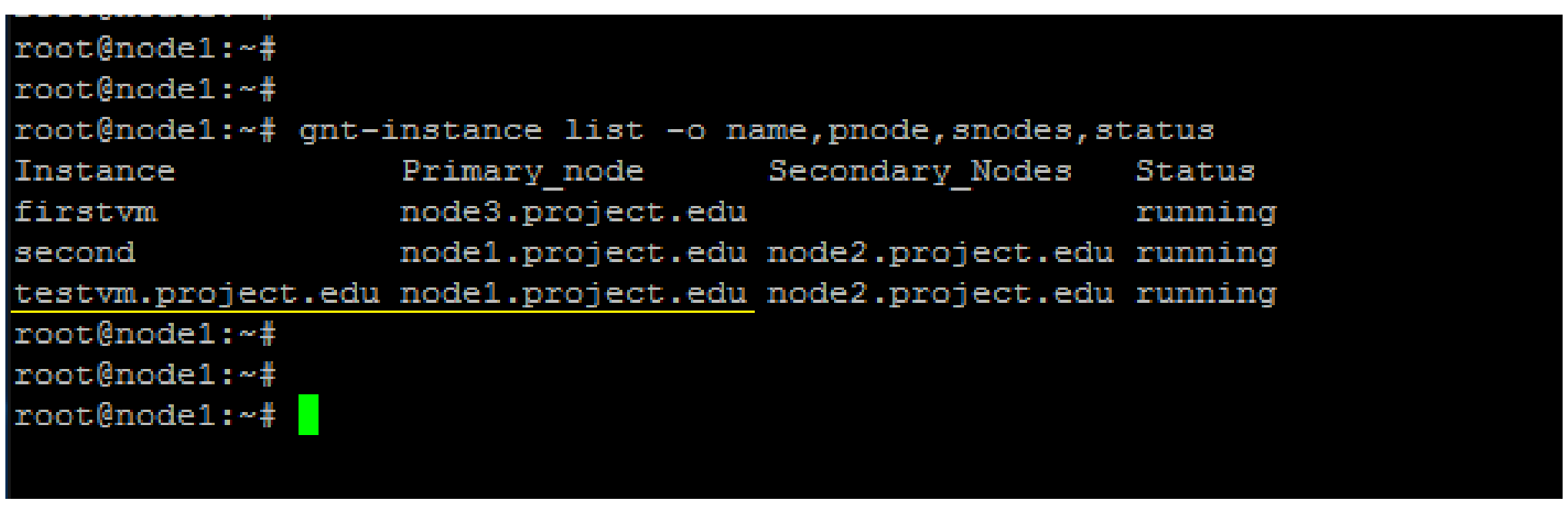

Figure 3.13: Check total instances of the Cluster

## 3. Conclusion and Future Work

In our project we have tried our best to run the virtualization over commodity hardware and create some VMs on it. We have successfully completed the job. Later we tried to introduce some scenarios and recommended some standard way-out of those cases. We can suggest this project for small and medium office if they want to move for virtualization of their services using existing commodity hardware.

Probing deeper, one can use a web management tool for Ganeti administration. Moreover if the cluster used for business and provided SaaS to the customers, one can work on the development of a web interface for system administrators so that they can manage and check billing of their uses which will be a very useful tool for provider as well as customer.

Table 1. Heading and text fonts.

| Text | Alignment | Font | Followed by: |
|---|---|---|---|
| Title | Centre | 20 pt. TNR, bold, small-caps | 24 pt. line sp. |
| Authors | Centre | 13 pt. TNR | 12 pt. line sp. |
| Addresses | Centre | 12 pt. TNR | |
| emails | Centre | 11 pt. italic TNR | 18 pt. line sp. (last) |
| Abstract heading | Left | 13 pt. bold italic TNR, small caps | 6 pt. line sp. |
| Abstract text | Left | 10 pt. italic TNR | 12 pt. line sp. |
| Keywords heading | Left | 13 pt. bold italic TNR, small caps | 6 pt. line sp. |
| Keywords | Left, left, .. | 10 pt. italic TNR | 18 pt line sp. |
| Section headings | Left | 14 pt. bold TNR, small caps | 6 pt. line sp. |
| Sub-section heads | Left | 12 pt. bold TNR | 6 pt. line sp. |
| Sub-sub-sections | Left | 11 pt. bold TNR | 6 pt. line sp. |
| Body text | Full (left/right) | 11 pt. TNR | 12 pt line sp. (last) |
| Figures | Centre | | 6 pt. line sp. |
| Figure captions | Centre | 11 pt. TNR | 12 pt. line sp. |
| References | Left | 10 pt. TNR (as shown) | 6 pt. line sp |

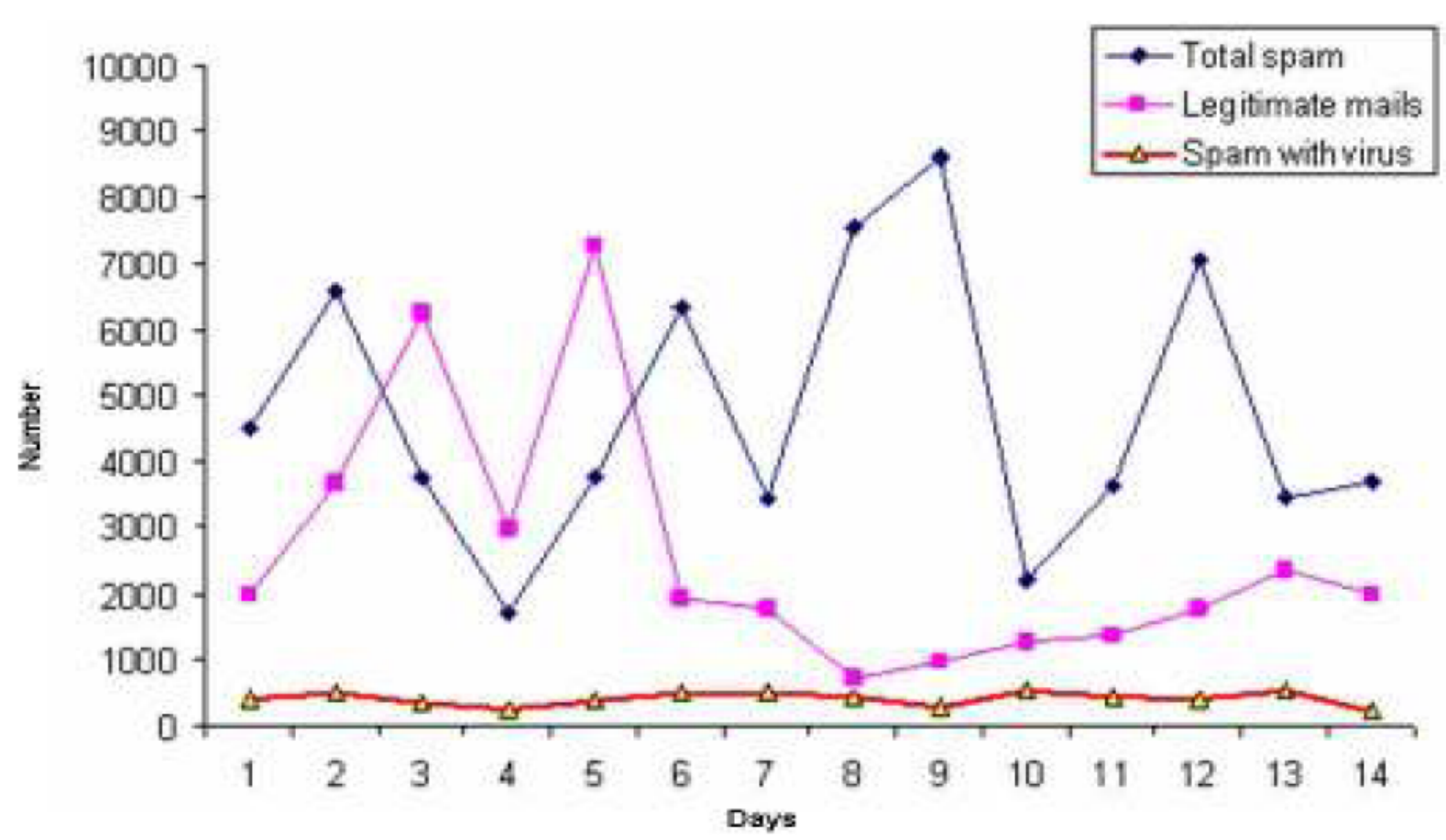

Figure 4. Spam traffic sample

## Authors

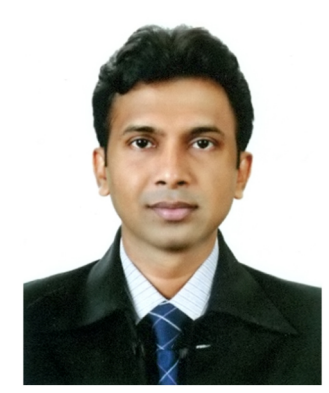

Mohammad Mamun Or Rashid received his B.Sc. (Hon’s) in Computer Science from North South University (NSU), Dhaka, Bangladesh in 2006 and M.Sc. in Computer Science in 2015 from Jahangirnagar University, Savar, Dhaka, Bangladesh. He has been working in Government of the People's Republic of Bangladesh as a “System Analyst” in Ministry of Expatriate’s Welfare and Overseas Employment. His current research interests include Cloud Computing, virtualization and information Security management system. He is also interested in Linux and Virtual networking in cloud computing.

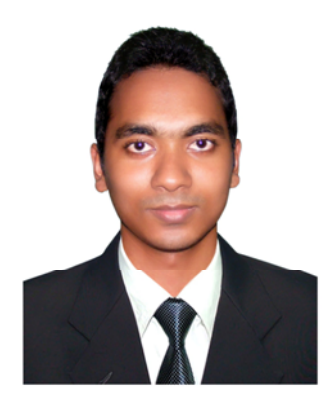

M. Masud Rana received the B.Sc. in Computer Science and Engineering from the Dhaka International University, Dhaka, Bangladesh in 2014. Currently, he is working towards M.Sc. in Computer Science from the Jahangirnagar University, Savar, Dhaka, Bangladesh. He has serving as an Executive Engineer, Information Technology in Bashundhara Group also he has more than 5 years of experience as an Assistant Engineer, IT in SQUARE Informatix Ltd, Bangladesh and Executive Engineer, IT in Computer Source Ltd, Bangladesh. His main areas of research interests include virtualization, networking and security aspects of cloud computing.

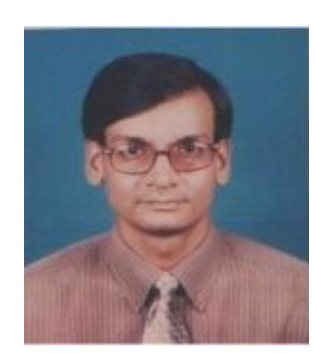

Jugal Krishna Das received B.Sc., M.Sc. and PhD in Computer Science all from Rassia. He is currently an Professor of Computer Science and Engineering department of Jahangirnagar University, Savar, Dhaka. His research interests include topics such as Computer Networks, Natural Language Processing, Software Engineering.